\documentclass[sigconf]{acmart}
\AtBeginDocument{%
  }

\setcopyright{acmlicensed}
\copyrightyear{2026}
\acmYear{2026}
\acmDOI{XXXXXXX.XXXXXXX}
\acmConference[KDD '26]{(32nd ACM SIGKDD Conference on Knowledge Discovery and Data Mining}{August 09--13,
  2026}{Jeju, Korea}
\usepackage{booktabs}   % For \toprule, \midrule, \bottomrule
\usepackage{multirow}   % For \multirow
\usepackage{graphicx}   % For \resizebox
\usepackage{color}    % For \textcolor
\usepackage{bm}         % For bold math symbols
\usepackage{subcaption}
\usepackage{balance}

\begin{document}

\title[PhyGHT for Signal Purification at the HL-LHC]{PhyGHT: Physics-Guided HyperGraph Transformer \\for Signal Purification at the HL-LHC}

\author{Mohammed Rakib}
\affiliation{%
\institution{Department of Computer Science}
\country{Oklahoma State University, USA}
}
\email{mohammed.rakib@okstate.edu}

\author{Luke Vaughan}
\affiliation{%
\institution{Department of Physics}
\country{Oklahoma State University, USA}
}
\email{luke.vaughan@okstate.edu}

\author{Shivang Patel}
\affiliation{%
\institution{Department of Physics}
\country{Oklahoma State University, USA}
}
\email{shivang.patel@okstate.edu}

\author{Flera Rizatdinova}
\affiliation{%
\institution{Department of Physics}
\country{Oklahoma State University, USA}
}
\email{flera.rizatdinova@okstate.edu}

\author{Alexander Khanov}
\affiliation{%
\institution{Department of Physics}
\country{Oklahoma State University, USA}
}
\email{alexander.khanov@okstate.edu}

\author{Atriya Sen}
\affiliation{%
\institution{Department of Computer Science}
\country{Oklahoma State University, USA}
}
\email{atriya.sen@okstate.edu}

\renewcommand{\shortauthors}{Rakib et al.}

% \begin{abstract}
%  The High-Luminosity Large Hadron Collider (HL-LHC) at CERN will produce unprecedented datasets capable of revealing fundamental properties of the universe. However, realizing its discovery potential faces a significant challenge: extracting small signal fractions from overwhelming backgrounds dominated by approximately 200 simultaneous pileup collisions. This extreme pileup creates a noisy environment that severely distorts physical observables. To tackle this challenge, we propose the \textit{Physics-Guided Hypergraph Transformer} (PhyGHT), a heterogeneous graph neural network that directly regresses signal observables from pileup-dominated data. PhyGHT utilizes a novel \textit{Pileup Suppression Gate} (PSG) and \textit{Hypergraph Attention} mechanism to dynamically fuse purified track features into robust jet representations. By creating a novel simulated dataset under extreme pileup conditions, PhyGHT directly predicts energy and mass correction factors for the signal and outperforms state-of-the-art methods from the ATLAS and CMS experiments. We demonstrate PhyGHT's real-world impact by accurately reconstructing the top quark mass, showing how physics-informed deep learning architectures can restore precision measurements that would otherwise be obscured by noise. This work exemplifies how machine learning innovation and interdisciplinary collaboration can directly advance scientific discovery at the frontiers of experimental physics and enhance the discovery potential of the HL-LHC.
% \end{abstract}

\begin{abstract}
The High-Luminosity Large Hadron Collider (HL-LHC) at CERN will produce unprecedented datasets capable of revealing fundamental properties of the universe. However, realizing its discovery potential faces a significant challenge: extracting small signal fractions from overwhelming backgrounds dominated by approximately 200 simultaneous pileup collisions. This extreme noise severely distorts the physical observables required for accurate reconstruction. To address this, we introduce the \textit{Physics-Guided Hypergraph Transformer} (PhyGHT), a hybrid architecture that combines \textit{distance-aware local graph attention} with \textit{global self-attention} to mirror the physical topology of particle showers formed in proton-proton collisions. Crucially, we integrate a \textit{Pileup Suppression Gate} (PSG), an interpretable, physics-constrained mechanism that explicitly learns to filter soft noise prior to hypergraph aggregation. To validate our approach, we release a novel simulated dataset of top-quark pair production to model extreme pileup conditions. PhyGHT outperforms state-of-the-art baselines from the ATLAS and CMS experiments in predicting the signal's energy and mass correction factors. By accurately reconstructing the top quark's invariant mass, we demonstrate how machine learning innovation and interdisciplinary collaboration can directly advance scientific discovery at the frontiers of experimental physics and enhance the HL-LHC's discovery potential. \textit{The dataset and code are available at https://github.com/rAIson-Lab/PhyGHT.}
\end{abstract}


\begin{teaserfigure}
  \includegraphics[width=\textwidth]{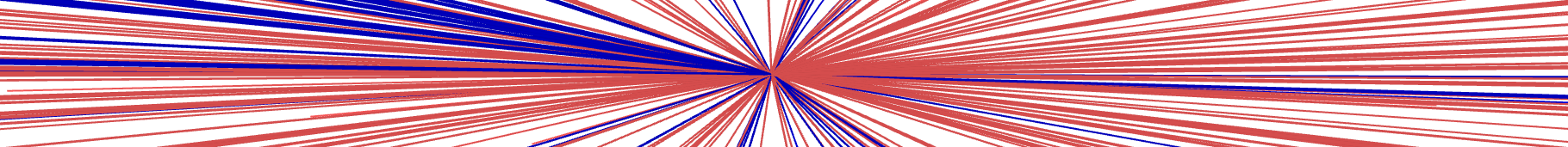}
  \caption{Signal Collision (Blue) covered in Pileup Collisions (Red) in the HL-LHC Environment}
  %\Description{Enjoying the baseball game from the third-base
  %seats. Ichiro Suzuki preparing to bat.}
  \label{fig:teaser}
\end{teaserfigure}

\received{20 February 2007}
\received[revised]{12 March 2009}
\received[accepted]{5 June 2009}

\maketitle
    
\section{Introduction}

%\begin{figure*}[t!]
%    \centering
%    \includegraphics[width=0.8\linewidth]{figures/lukes_upload/Pileup_Graphic.png}
%    \caption{Schematic Event Display: NOT FINAL VERSIONS}
%    \label{fig:EventDisplay}
%\end{figure*}

\begin{figure*}[t!]
    \centering
    \begin{subfigure}{.49\textwidth}
      \centering
      \includegraphics[width=0.82\linewidth]{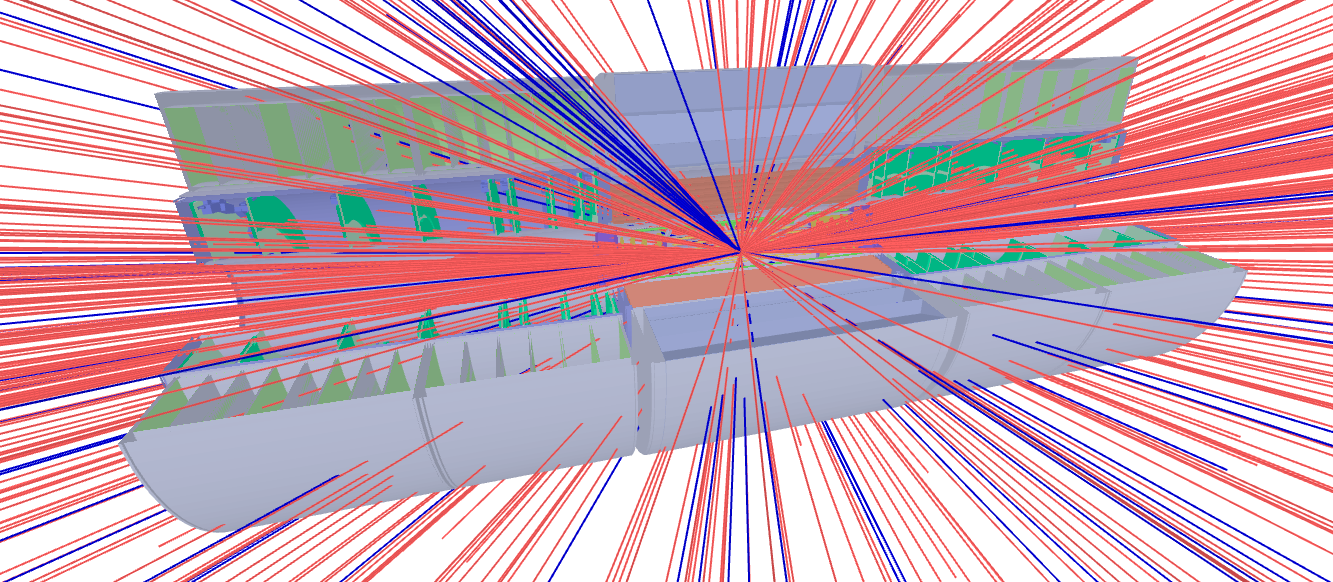}
      \caption{}
      \label{fig:Efrac1d_mu60}
    \end{subfigure}\hfill
    \begin{subfigure}{.49\textwidth}
      \centering
      \includegraphics[width=0.82\linewidth]{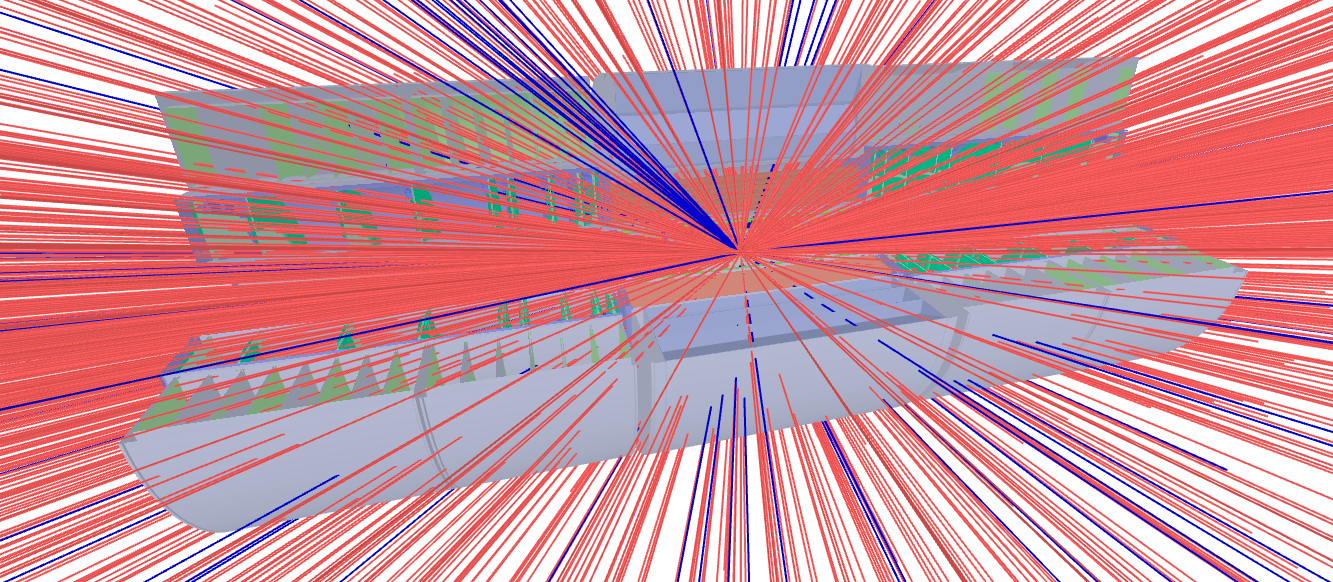}
      \caption{}
      \label{fig:Efrac2d_mu60}
    \end{subfigure}\hfill
    \caption{A Phoenix Event Display\cite{phoenix} depicting an event in the ATLAS inner tracking system in the HL-LHC conditions at $\langle\mu\rangle=60$ (left) and $\langle\mu\rangle=200$ (right) for signal (blue) and background (red) particles with $p_T>1.0$ GeV.}
    \label{fig:EventDisplay}
\end{figure*}

Starting in 2030, the Large Hadron Collider (LHC) at CERN will begin upgrades to the High-Luminosity phase (HL-LHC)\cite{HL-LHC}. High-energy physics experiments, such as ATLAS\cite{ATLAS} and CMS\cite{CMS}, will record vast amounts of data by colliding bunches of protons to better understand the fundamental structure of nature.
%When a collision satisfies the preselection filter criteria (a trigger), an \textit{event} is recorded consisting of both the particles produced in that collision (signal) and particles from all other collisions occurring within the same bunch crossing (background). The background collisions from each proton bunch crossing are referred to as \textit{pileup}. At the HL-LHC, the average number of pileup interactions, denoted as $\langle\mu\rangle$, will increase up to 200 as shown in Fig \ref{fig:EventDisplay}.
Since it is not feasible to collect data from all collisions, only "interesting" \textit{signal} interactions that satisfy online preselection criteria will be recorded. Interactions recorded by the detector data acquisition system, called \textit{events}, will include particles due to both the signal interaction and additional spurious interactions within the same bunch crossing, referred to as \textit{pileup}. The average number of interactions per bunch crossing, defined as $\langle\mu\rangle$, will shift from 60 at the LHC to 200 at the HL-LHC environment, which will significantly increase the pileup background, as shown in Fig.~\ref{fig:EventDisplay}.

%In many cases, particles produced in these collisions will form collimated streams called \textit{jets}. The ATLAS and CMS experiments use many detector subsystems to reconstruct jets. In this paper, we construct jets purely as a set of particles. Jets that contain significant contributions from the signal collision are of interest to physics analysis; however, signal jets have mass and energy distorted by the pileup background. The task of \textit{pileup mitigation} attempts to suppress the pileup background and restore the physical quantities from the signal collision.
Due to the nature of physics processes at the LHC, particles created during the proton collisions often form collimated streams called \textit{jets}, which are observed as compact clusters of energy in the detector. In the presence of pileup, the energy and momentum of signal jets become significantly distorted compared to their true values, which hinders physics analysis. The task of pileup mitigation is to suppress the pileup background and restore the physical quantities from the signal collision.

While the HL-LHC will maximize the potential for discoveries in particle physics, the unprecedented levels of pileup will pose a major challenge for data analysis. Finding the signal of a rare physics process in this environment is akin to finding a needle in a haystack datasets \cite{trackml, lhcolympic}, where machine learning (ML) models must be robust enough to denoise complex event data. In recent years, there is a rising interest in ML algorithms for mitigating pileup effects\cite{pumml,Maier_2022,Mikuni_2020,martinez2019,carlson2025}. Current non-ML production algorithms deployed in LHC experiments, typically optimized for lower pileup conditions, have adopted two distinct approaches: either mitigating pileup at the jet level \cite{jvt} or at the particle level \cite{puppi}. These single-modality strategies often face a trade-off: jet-level substructure can obscure internal substructure, while particle-level filtering may lack the global context required to estimate event-wide noise density. In our approach, we exploit the full information from both particles and jets in each event by leveraging a combination of graph architectures \cite{martinez2019,particlenet,GN2,hgpflow} and transformer encoders \cite{part,Maier_2022,puminet} to directly provide energy and mass correction factors for each jet, enabling a more precise pileup mitigation.

In this work, we propose the Physics-Guided Hypergraph Transformer (PhyGHT), a hierarchical architecture that fuses Distance-Aware Graph Attention (DA-GAT) for local sub-structure encoding with a Global Transformer for event-level energy flow analysis. This hybrid approach resolves the trade-off between local geometric precision and global context awareness required for high-pileup environments. Inspired by PUPPI \cite{puppi}, we introduce the Pileup Suppression Gate (PSG), a learnable and differentiable mechanism designed to enhance interpretability. PSG explicitly predicts a per-particle signal probability, enabling the model to perform soft-masking of pileup prior to aggregation with jets
We formulate the jet purification task as a hypergraph aggregation problem, treating jets as hyperedges that connect variable-sized sets of tracks. This approach employs a bipartite attention mechanism to dynamically weight constituent tracks, overcoming the information loss associated with fixed-size pooling \cite{gnn} and enabling precise reconstruction of physical observables such as top quark mass. To rigorously evaluate these capabilities, we release a simulated dataset of top quarks under extreme pileup conditions ($\langle\mu\rangle=200$) that closely mimic the real-world detector data.

%Our primary contributions are as follows:
%\begin{itemize}
%    \item We propose the Physics-Guided Hypergraph Transformer (PhyGHT), a novel heterogeneous architecture that fuses Distance-Aware Graph Attention (DA-GAT) for local substructure encoding with a Global Transformer for event-wide energy flow analysis. This hybrid approach resolves the trade-off between local geometric precision and global context awareness required for high-pileup environments.
%    \item Inspired by PUPPI \cite{puppi}, we introduce the Pileup Suppression Gate (PSG), a learnable and differentiable mechanism designed to enhance interpretability. PSG explicitly predicts a per-particle signal probability, enabling the model to perform soft-masking of pileup prior to aggregation with jets
%    \item We formulate the jet pruning task as a hypergraph aggregation problem, treating jets as hyperedges that connect variable-sized sets of tracks. This approach employs a bipartite attention mechanism to dynamically weight constituent tracks, overcoming the information loss associated with fixed-size pooling \cite{GraphSage} and enabling precise recovery of sensitive substructure observables, such as Jet Mass.
%\end{itemize}
\section{Dataset and Simulation}
\label{sec:dataset}
We present a novel, open-source dataset that fills a critical need at the intersection of machine learning and high-energy physics. While functionally similar to proprietary datasets used by the ATLAS collaboration~\cite{ANA-FTAG-2024-06}, our dataset offers three key advantages. First, it provides truth labels for pileup characterization. This enables precise separation of signal and background noise under HL-LHC conditions. Second, it is fully public. This provides open access to realistic particle physics data that is typically restricted to collaboration members. Third, it supports reproducibility and scalability. This allows the broader computer science community to comprehensively evaluate state-of-the-art machine learning architectures on complex particle physics analysis tasks. By simulating the extreme pileup conditions ($\langle\mu\rangle=200$) expected at the HL-LHC, we provide a benchmark that allows researchers to build and test systems ready for the next generation of particle physics experiments. The dataset is available at \href{https://zenodo.org/records/18746388}{Zenodo}.

\subsection{Data Representation}
The signal process is chosen to be top quark pair production decaying semi-leptonically, ($pp\to t\bar{t}, t \to q\bar{q}'b, \bar{t} \to \ell\nu\bar{b}$), and is generated using \textit{MadGraph5\_aMC@NLO} \cite{madgraph}. \textit{Pythia 8} \cite{pythia} with ATLAS A14 central tune\cite{atlas_tune,nnpdf} is used for parton showering which results in stable, final state particles which can be observed by the detectors. To mimic both standard LHC conditions and the high pileup ones of HL-LHC, soft Quantum Chromodynamics (QCD) pileup interactions were overlaid by sampling from a Poisson distribution with mean $\langle\mu\rangle = 60$ and $\langle\mu\rangle = 200$, respectively. The primary and pileup vertices were spatially smeared using Gaussian distributions with widths $\sigma_{xy} = 0.3$ mm and $\sigma_{z} = 50$ mm. Stable, final state particles are clustered using \textit{FastJet} \cite{fastjet} with the anti-$k_t$ algorithm \cite{antikt} using a cone size parameter of $R=0.4$ and the minimum transverse momentum threshold $p_T^{min}>25GeV$. Lastly to model detector acceptance, we removed neutral particles and charged particles with $p_T < 400$ MeV. 

Particles in the detector are described using a standard 3D coordinate system defined by their transverse momentum ($p_T$), pseudorapidity ($\eta$), and azimuthal angle ($\phi$)\footnote{$\phi$ is azimuthal angle, $\eta=-\ln\left[ \tan\left( \frac{\theta}{2} \right) \right]$ where $\theta$ is polar angle, and $p_{\rm T}=|\overrightarrow{p}|\sin\theta$ in the spherical coordinate system where the $z$ axis is directed along the beam.}. Each charged particle track is represented by a feature vector $\mathbf{x}^{track} = [p_T, \eta, \phi, q, d_0, z_0]$. Here, $q$ is the charge, and $d_0$ and $z_0$ denote the transverse and longitudinal impact parameters, respectively. These impact parameters are calculated by extrapolating the track to the beam line \cite{beamline}. Each jet is described by a vector $\mathbf{x}^{jet} = [p_T, \eta, \phi, m]$ where $m$ is the mass. These features of the jet represent the aggregated kinematics of the charged and neutral constituents of each clustered set of particles.

\subsection{Truth Label Definition}
Each jet and track are described as a Lorentz 4-vector which is defined by energy and momentum: $(E, \vec{p})$. The mass of an object can be calculated using the relativistic energy-momentum relation as $m=\sqrt{E^2-|\vec{p}|^2}$. Each jet is constructed by summing the 4-vectors over a set of tracks. Each track is assigned a binary label $y^{label} \in \{0, 1\}$. We assign a value of 1 to tracks originating from the signal vertex and 0 to those originating from pileup vertices, which are used as an auxillary task during training. For each jet, we can calculate truth level energy correction factor $y_{E, k}$ and mass correction factor $y_{M, k}$ for the $k$-th jet as the ratio of the contributions from HS tracks to the total contributions from all tracks:
\begin{equation}
    y_{E, k} = \frac{E_{HS, k}}{E_{raw, k}}, \quad y_{M, k} = \frac{M_{HS, k}}{M_{raw, k}}
    \label{eq:truth_labels}
\end{equation}
where $E_{raw, k}$ and $M_{raw, k}$ represent the total jet energy and mass clustered from the hard scatter signal and the pileup background.
\section{Methodology}
\label{sec:methodology}

\begin{figure*}[t]
  \centering
  \includegraphics[width=\textwidth]{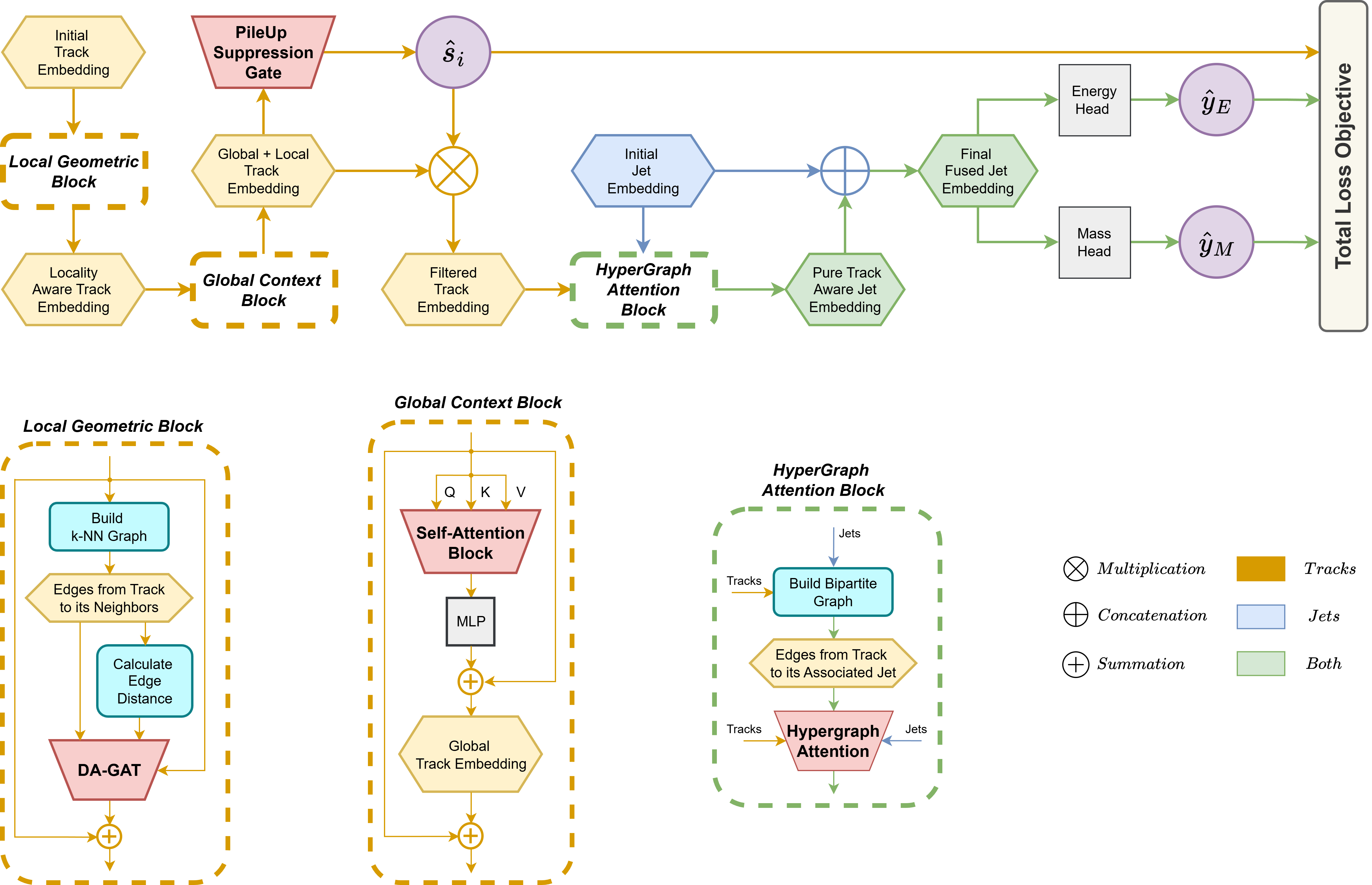}
  \caption{Overview of PhyGHT. It takes a heterogeneous graph of Tracks and Jets as input. It processes tracks through the Local Geometric (DA-GAT) block, followed by the Global Context block. The fused representations are filtered by the Pileup Suppression Gate (PSG) before being aggregated into the Jet representation via Hypergraph Attention for final regression.}
  \label{fig:architecture}
\end{figure*}

% Figure~\ref{fig:architecture} illustrates our PhyGHT architecture, a hierarchical graph neural network designed to mitigate pileup-induced distortion in physical jet observables such as mass and energy. PhyGHT takes as input raw event-level information, specifically jets and tracks. It then employs a sequential feature refinement strategy to purify jets from pileup. The process begins with the \textit{Local Geometric Block}, which encodes the collinear substructure of track clusters using a Distance-Aware Graph Attention Network (DA-GAT). This locality-aware representation is then refined by the \textit{Global Context Block}, where a Transformer encoder captures event-wide correlations such as pileup density. To ensure physical robustness, the \textit{Pileup Suppression Gate (PSG)} applies a differentiable \textit{soft-mask} filter to these fused features, explicitly suppressing noise prior to aggregation. Finally, the \textit{Hypergraph Attention Block} fuses the purified tracks with raw jet features via bipartite message passing, enabling the precise regression of energy and mass correction factors.

Figure~\ref{fig:architecture} illustrates our PhyGHT architecture, a hierarchical graph neural network designed to mitigate pileup-induced distortion in physical jet observables such as mass and energy. PhyGHT takes as input raw event-level information, specifically jets and tracks. It then employs a \textit{four-stage} sequential feature refinement strategy to purify jets from pileup. First, the \textit{Local Geometric Block} uses a Distance-Aware Graph Attention Network (DA-GAT) to encode spatial correlations, effectively capturing the local topology of signal particles against stochastic pileup. Then the \textit{Global Context Block} employs a Transformer encoder to model event-wide constraints like pileup density and momentum conservation. Next, the \textit{Pileup Suppression Gate (PSG)} applies a learnable, differentiable mask to down-weight pileup tracks, serving as an analogue to traditional algorithms such as PUPPI. Finally, the \textit{Hypergraph Attention Block} dynamically aggregates purified tracks and fuses them with raw jet features via bipartite message passing, enabling precise regression of energy and mass correction factors.

\subsection{Problem Formulation}
\label{sec:problem_formulation}

We formulate pileup mitigation as a regression task on a heterogeneous graph, aiming to predict correction factors that recover hard-scatter observables from pileup-contaminated events. Let a collision event be represented as a graph $\mathcal{G} = (\mathcal{V}, \mathcal{E})$, where $\mathcal{V} = \mathcal{V}_{track} \cup \mathcal{V}_{jet}$ consists of $N_{track}$ track nodes and $N_{jet}$ jet nodes. We denote track indices by $i, j \in \{1, \dots, N_{track}\}$ and jet indices by $k \in \{1, \dots, N_{jet}\}$.

\subsubsection{Node Features}
Each track node $t_i \in \mathcal{V}_{track}$ is initialized with a feature vector $\mathbf{x}_i^{track} \in \mathbb{R}^{6}$ containing the kinematic and vertexing variables defined in Section~\ref{sec:dataset}. Similarly, each jet node $j_k \in \mathcal{V}_{jet}$ acts as a hypernode initialized with the aggregate feature vector $\mathbf{x}_k^{jet} \in \mathbb{R}^{4}$. These features represent the raw input state.
% prior to any correction

\subsubsection{Graph Connectivity}
The edge set $\mathcal{E}$ encodes both local geometric and hierarchical relationships through two distinct subsets. First, the \textit{Local Edge Set} $\mathcal{E}_{local}$ connects track nodes to their spatial neighbors; an edge $(t_i, t_j) \in \mathcal{E}_{local}$ exists if track $t_j$ is among the $K$-nearest neighbors of track $t_i$ in the $(\eta, \phi)$ metric space, defined by the Euclidean distance $\Delta R_{ij} = \sqrt{(\Delta\eta_{ij})^2 + (\Delta\phi_{ij})^2}$. Second, the \textit{Hypergraph Edge Set} $\mathcal{E}_{hyper}$ connects tracks to jets based on the clustering history, where a directed edge $(t_i, j_k)$ exists if track $t_i$ is a physical constituent of jet $j_k$.

\subsubsection{Learning Objective}
The goal of PhyGHT is to learn a mapping $f_\theta(\mathcal{G}) \to \hat{\mathbf{Y}}$ that regresses the hard-scatter contributions. For each jet $j_k$, the model predicts an estimated energy correction factor $\hat{y}_{E, k} \in [0, 1]$ and mass correction factor $\hat{y}_{M, k} \in [0, 1]$ corresponding to the ground truth ratios in Eq.~\ref{eq:truth_labels}. Targeting these bounded coefficients rather than absolute values stabilizes the regression against the large dynamic range of jet kinematics. The final physical quantities are reconstructed via:
\begin{equation}
    E_{corr, k} = \hat{y}_{E, k} \cdot E_{raw, k}, \quad M_{corr, k} = \hat{y}_{M, k} \cdot M_{raw, k}
\end{equation}
where $E_{corr, k}$ and $M_{corr, k}$ represent the corrected energy and mass.
% of the raw jet. 

\subsection{Local Geometric Encoding}
\label{sec:local_encoding}

The primary objective of the Local Geometric block is to encode the local context of particle showers represented by jets. In high-energy physics, signal particles from a hard scatter typically exhibit collinearity, clustering tightly in the $(\eta, \phi)$ space, whereas particles from 200 superimposed pileup collision create a random distribution of tracks. To capture this, we employ a Distance-Aware Graph Attention Network (DA-GAT) that biases the aggregation of neighbor features based on their spatial proximity in the detector.

\subsubsection{Feature Embedding}
First, the raw input features $\mathbf{x}_i^{track}$ of each track $t_i$ are projected into a high-dimensional latent space to enable the learning of non-linear kinematic correlations. We apply a linear transformation followed by layer normalization and a GELU activation to get the initial track embedding $\mathbf{h}_i^{(0)} \in \mathbb{R}^{D}$:
\begin{equation}
    \mathbf{h}_i^{(0)} = \text{GELU}\left( \text{LayerNorm}\left( \mathbf{W}_{T} \mathbf{x}_i^{track} + \mathbf{b}_{T} \right) \right)
\end{equation}
where $\mathbf{W}_{T} \in \mathbb{R}^{D \times 6}$, $\mathbf{b}_{T} \in \mathbb{R}^{D}$ and $D$ is the hidden dimension.

\subsubsection{Distance-Aware Graph Attention (DA-GAT)}
Standard Graph Attention Networks (GATs) \cite{gat} compute attention coefficients solely based on node features, potentially assigning high weights to distant pileup tracks that share similar kinematic properties with the signal. To mitigate this, we inject a structural bias into the attention mechanism. For every track $t_i$, we consider its local neighborhood $\mathcal{N}_K(i)$ defined by $k$-NN in the $\mathcal{E}_{local}$ edge set. We compute the normalized attention coefficients $\alpha_{ij}$ directly via:
\begin{equation}
    \alpha_{ij} = \mathcal{S}\left( \text{LeakyReLU}\left( \mathbf{a}^T \left[ \mathbf{W} \mathbf{h}_i^{(0)} \, \big\Vert \, \mathbf{W} \mathbf{h}_j^{(0)} \, \big\Vert \, w_d \cdot (\Delta R_{ij})^2 \right] \right) \right)
\end{equation}
Here, $\Vert$ denotes concatenation, $\mathbf{W} \in \mathbb{R}^{D \times D}$ is a shared weight matrix, $\mathbf{a} \in \mathbb{R}^{2D+1}$ is the attention vector, $\Delta R_{ij}$ is the Euclidean distance in the $(\eta, \phi)$ plane, and $\mathcal{S}$ denotes the softmax function applied across the neighborhood $\mathcal{N}_K(i)$. Crucially, $w_d$ is a learnable scalar parameter. This allows the network to learn a spatial decay function analogous to a Gaussian kernel that penalizes information flow from physically distant tracks, effectively enforcing a \textit{soft cone} size for information aggregation.

\subsubsection{Aggregation}
The local representation for track $t_i$ is updated by aggregating the neighbor features weighted by $\alpha_{ij}$: 
\begin{equation}
    \mathbf{h}_i^{(agg)} = \sum_{j \in \mathcal{N}_K(i)} \alpha_{ij} \mathbf{W} \mathbf{h}_j^{(0)}
\end{equation}
To preserve gradient flow and stabilize training, we employ a residual connection and layer normalization:
\begin{equation}
    \mathbf{h}_i^{(local)} = \text{LayerNorm}\left( \mathbf{h}_i^{(0)} + \text{Dropout}\left( \text{GELU}\left( \mathbf{h}_i^{(agg)} \right) \right) \right)
\end{equation}
The resulting vector $\mathbf{h}_i^{(local)}$ encodes the track's kinematic state contextually enriched by its immediate geometric surroundings.

\subsection{Global Contextualization}
\label{sec:global_context}

While DA-GAT captures the fine-grained local context of jets, it is inherently blind to long-range dependencies such as global momentum conservation and event-wide pileup density fluctuations. To address this, we process the locally encoded features $\mathbf{h}_i^{(local)}$ through a Global Contextualization block based on the Transformer encoder architecture \cite{transformer}.

\subsubsection{Global Self-Attention}
We treat the event as a fully connected graph where every track attends to every other track. For each track $t_i$, we compute Query ($\mathbf{q}_i$), Key ($\mathbf{k}_i$), and Value ($\mathbf{v}_i$) vectors via linear projections below, where $\mathbf{W}_Q, \mathbf{W}_K, \mathbf{W}_V \in \mathbb{R}^{D \times D}$:
\begin{equation}
    \mathbf{q}_i = \mathbf{W}_Q \mathbf{h}_i^{(local)}, \quad \mathbf{k}_i = \mathbf{W}_K \mathbf{h}_i^{(local)}, \quad \mathbf{v}_i = \mathbf{W}_V \mathbf{h}_i^{(local)}
\end{equation}

The pairwise attention coefficients $A_{ij}$, representing the relevance of track $t_j$ to track $t_i$, are computed via the scaled dot-product:
\begin{equation}
    A_{ij} = \mathcal{S} \left( \frac{\mathbf{q}_i^T \mathbf{k}_j}{\sqrt{D}} \right)
\end{equation}
where $\mathcal{S}$ denotes the softmax function applied across all tracks $j \in \mathcal{V}_{track}$. The global context vector $\mathbf{h}_i^{(global)} \in \mathbb{R}^{D}$ is obtained by aggregating the values weighted by the attention coefficients, followed by a Feed-Forward Network (FFN), residual connection, and layer normalization:
\begin{equation}
    \mathbf{h}_i^{(global)} = \text{LayerNorm}\left( \mathbf{h}_i^{(local)} + \text{FFN}\left( \sum_{j=1}^{N_{track}} A_{ij} \mathbf{v}_j \right) \right)
\end{equation}

\subsubsection{Glocal Fusion}
 Finally, to preserve the strong geometric gradients learned by the DA-GAT, we fuse the global and local representations through summation:
\begin{equation}
    \mathbf{z}_i = \mathbf{h}_i^{(local)} + \mathbf{h}_i^{(global)}
\end{equation}
The resulting vector $\mathbf{z}_i \in \mathbb{R}^{D}$ serves as the input to the gating mechanism, encoding both the dense collinear structure of the jet and the global event context.

\subsection{Pileup Suppression Gate (PSG)}
\label{sec:psg}

We introduce the Pileup Suppression Gate (PSG), a learnable mechanism inspired by the phenomenological PUPPI algorithm \cite{puppi}. While standard attention mechanisms implicitly down-weight noisy features, they do not explicitly remove them. The PSG acts as a differentiable \textit{soft-mask} filter, explicitly predicting the probability that a track originates from the hard-scatter vertex.

\subsubsection{Signal Probability Estimation}
The fused feature vector $\mathbf{z}_i$, is passed through a Multi-Layer Perceptron (MLP) to compute a scalar signal probability score $\hat{s}_i \in [0, 1]$:
\begin{equation}
    \hat{s}_i = \sigma\left( \mathbf{w}_{gate}^T \cdot \text{Dropout}\left( \text{ReLU}\left( \mathbf{W}_{gate} \mathbf{z}_i + \mathbf{b}_{gate} \right) \right) \right)
\end{equation}
where $\mathbf{W}_{gate} \in \mathbb{R}^{d_{gate} \times D}$, $\mathbf{w}_{gate} \in \mathbb{R}^{d_{gate}}$, $\mathbf{b}_{gate} \in \mathbb{R}^{D}$ , and $\sigma$ is the sigmoid function. This score $\hat{s}_i$ serves as a track-level confidence metric, providing interpretability: values near 1 indicate signal-like kinematics, while values near 0 indicate pileup.

\subsubsection{Differentiable Filtering}
We apply this score to the feature vector via element-wise multiplication, effectively suppressing the influence of identified pileup tracks before aggregation:
\begin{equation}
    \tilde{\mathbf{z}}_i = \hat{s}_i \cdot \mathbf{z}_i
\end{equation}
This operation retains the feature direction for signal tracks while shrinking pileup vectors toward zero, effectively mitigating pileup.

\subsection{Hypergraph Attention Aggregation}
\label{sec:hypergraph}

We purify jets by aggregating the filtered constituent tracks via a hypergraph attention mechanism. This allows the model to dynamically weight tracks based on their relevance to the specific jet cluster, rather than relying on a fixed global pooling.

\subsubsection{Feature Projection}
We project the raw features of jets and the filtered features of tracks into a shared latent space $\mathbb{R}^D$. For each jet node $j_k$, we compute a query embedding $\mathbf{h}_k^{J}$ below from its initial features $\mathbf{x}_k^{jet} \in \mathbb{R}^4$, where $\mathbf{W}_{J} \in \mathbb{R}^{D \times 4}$ and $\mathbf{b}_{J} \in \mathbb{R}^{D}$:
\begin{equation}
    \mathbf{h}_k^{J} = \text{GELU}\left( \text{LayerNorm}\left( \mathbf{W}_{J} \mathbf{x}_k^{jet} + \mathbf{b}_{J} \right) \right)
\end{equation}
Simultaneously, for each track $t_i$, we project its filtered feature vector $\tilde{\mathbf{z}}_i \in \mathbb{R}^D$ to obtain a key embedding $\mathbf{h}_i^{T}$ below, where $\mathbf{W}_{T'} \in \mathbb{R}^{D \times D}$ and $\mathbf{b}_{T'} \in \mathbb{R}^{D}$:
\begin{equation}
    \mathbf{h}_i^{T} = \text{GELU}\left( \mathbf{W}_{T'} \tilde{\mathbf{z}}_i + \mathbf{b}_{T'} \right)
\end{equation}

\subsubsection{Bipartite Attention Mechanism}
We model aggregation as message passing on a bipartite graph where edges flow from constituent tracks to their parent jets. For a given jet $j_k$ and its constituent set $\mathcal{N}(j_k) = \{t_i \mid (t_i, j_k) \in \mathcal{E}_{hyper}\}$, we compute the attention coefficients $\beta_{ki}$ using a dynamic graph attention mechanism:
\begin{equation}
    \beta_{ki} = \mathcal{S} \left( \text{LeakyReLU}\left( \mathbf{a}^T \left[ \mathbf{h}_k^{J} \, \Vert \, \mathbf{h}_i^{T} \right] \right) \right)
\end{equation}
where $\mathbf{a} \in \mathbb{R}^{2D}$ is the attention vector, and $\mathcal{S}$ is the softmax function that normalizes scores across the constituent set $t_i \in \mathcal{N}(j_k)$. The aggregated jet representation is then computed as the weighted sum of the constituent embeddings:
\begin{equation}
\mathbf{h}_k^{agg} = \sum_{t_i \in \mathcal{N}(j_k)} \beta_{ki} \cdot \mathbf{h}_i^{T}
\end{equation}

\subsubsection{Final Jet Embedding}
To stabilize the regression, we fuse the \textit{cleaned} information $\mathbf{h}_k^{agg}$ with the original raw jet embedding $\mathbf{h}_k^{J}$, which serves as a stable prior for the total energy scale. The final jet representation $\mathbf{h}_k^{final}$ is obtained via concatenation and non-linear projection below, where $\mathbf{W}_{fuse} \in \mathbb{R}^{D \times 2D}$:
\begin{equation}
    \mathbf{h}_k^{final} = \text{Dropout}\left( \text{ReLU}\left( \mathbf{W}_{fuse} \left[ \mathbf{h}_k^{J} \, \Vert \, \mathbf{h}_k^{agg} \right] \right) \right)
\end{equation}
% where $\mathbf{W}_{fuse} \in \mathbb{R}^{D \times 2D}$. 

\subsection{Task-Specific Prediction Heads}
The final jet embedding $\mathbf{h}_k^{final}$ encodes both the stable global energy scale and the cleaned local jet context. To decouple the tasks of energy and mass reconstruction, we pass this shared representation through two separate Multi-Layer Perceptrons (MLP), denoted as $\Phi_E$ and $\Phi_M$. Each head projects the latent vector to a scalar correction factor, constrained to the range $[0, 1]$ via sigmoid ($\sigma$):
\begin{equation}
    \hat{y}_{E, k} = \sigma\left( \Phi_E( \mathbf{h}_k^{final} ) \right), \quad \hat{y}_{M, k} = \sigma\left( \Phi_M( \mathbf{h}_k^{final} ) \right)
\end{equation}
$\hat{y}_{E, k}$ and $\hat{y}_{M, k}$ represent the estimated fraction of the raw jet's energy and mass, attributable to the hard scatter interaction, respectively.

\subsection{Joint Learning Objective}
\label{sec:loss}

To enforce both accurate jet purification and accurate track classification, we train PhyGHT with a multi-task objective. The total loss $\mathcal{L}_{total}$ is a weighted sum of the primary regression loss and an auxiliary classification loss:
\begin{equation}
    \mathcal{L}_{total} = \mathcal{L}_{reg} + \lambda \mathcal{L}_{aux}
\end{equation}
where $\lambda$ controls the influence of the physics-guided supervision.

\subsubsection{Regression Loss}
The primary objective is to minimize the error in the predicted correction factors for all jets in the event. We employ the Mean Squared Error (MSE) between the predicted fractions $(\hat{y}_{E, k}, \hat{y}_{M, k})$ and the ground truth ratios $(y_{E, k}, y_{M, k})$:
\begin{equation}
    \mathcal{L}_{reg} = \frac{1}{N_{jet}} \sum_{k=1}^{N_{jet}} \left[ (\hat{y}_{E, k} - y_{E, k})^2 + (\hat{y}_{M, k} - y_{M, k})^2 \right]
\end{equation}

\subsubsection{Auxiliary Classification Loss}
To ensure PSG learns to correctly identify signal particles, we use Binary Cross-Entropy loss on the track-level scores, $\hat{s}_i$. This forces the latent \textit{soft-mask} mechanism to align with the true physical vertex association $y_i^{label} \in \{0, 1\}$:
\begin{equation}
    \mathcal{L}_{aux} = -\frac{1}{N_{track}} \sum_{i=1}^{N_{track}} \left[ y_i^{label} \log(\hat{s}_i) + (1 - y_i^{label}) \log(1 - \hat{s}_i) \right]
\end{equation}
By jointly optimizing $\mathcal{L}_{aux}$ and $\mathcal{L}_{reg}$, the model learns to filter pileup explicitly while refining the aggregated jet properties, resulting in physical corrections to jet energy and mass.
\begin{table*}
\centering
\caption{Performance Comparison on the Test Set. We report the Coefficient of Determination ($R^2$) for the energy ($\hat{y}_E$) and mass ($\hat{y}_M$) correction factors across two pileup scenarios: $\langle\mu\rangle \in \{60, 200\}$.}
\label{tab:main_results}
\resizebox{0.78\linewidth}{!}{%
\begin{tabular}{c|c|cccccccc|c} 
\toprule
\multirow{2}{*}{$\bm{\langle\mu\rangle}$} & \multirow{2}{*}{\textbf{Target}} & \multicolumn{8}{c|}{\textit{Baselines}}                                                                                                       & \textit{Ours}    \\
                                          &                                  & \textbf{Transformer} & \textbf{GNN} & \textbf{HGNN} & \textbf{GAT} & \textbf{HGAT} & \textbf{PUPPI} & \textbf{ParticleNet} & \textbf{PUMINet} & \textbf{PhyGHT}  \\ 
\cmidrule(lr){1-11}
\multirow{2}{*}{60}                       & \textit{Energy}                  & 0.812                & 0.841        & 0.821         & 0.839        & 0.792         & 0.769             & 0.869                & \underline{0.934}            & \textbf{0.943}   \\
                                          & \textit{Mass}                    & 0.634                & 0.694        & 0.662         & 0.683        & 0.621         & 0.549             & 0.748                & \underline{0.838}            & \textbf{0.869}   \\ 
\cmidrule(lr){1-11}
\multirow{2}{*}{200}                      & \textit{Energy}                  & 0.778                & 0.837        & 0.798         & 0.792        & 0.756         & 0.348          & 0.853                & \underline{0.926}            & \textbf{0.932}   \\
                                          & \textit{Mass}                    & 0.567                & 0.643        & 0.587         & 0.591        & 0.595         & 0.114          & 0.693                & \underline{0.805}            & \textbf{0.836}   \\
\bottomrule
\end{tabular}
}%
\end{table*}

\begin{figure*}
    \centering
    % --- Row 1: mu = 60 ---
    \begin{subfigure}{.20\textwidth}
      \centering
      \includegraphics[width=\linewidth]{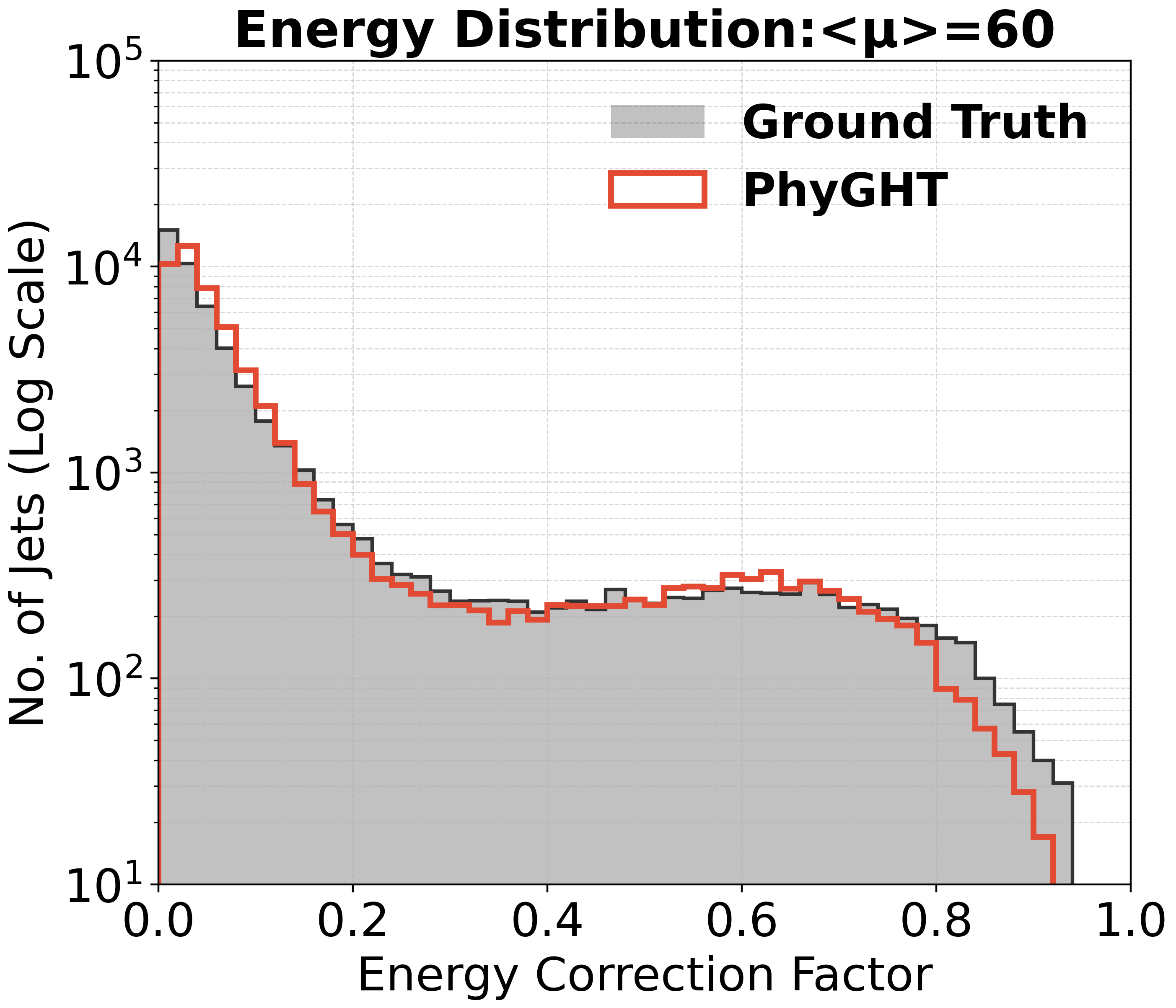}
      \caption{}
      \label{fig:Efrac1d_mu60}
    \end{subfigure}\hfill
    \begin{subfigure}{.20\textwidth}
      \centering
      \includegraphics[width=\linewidth]{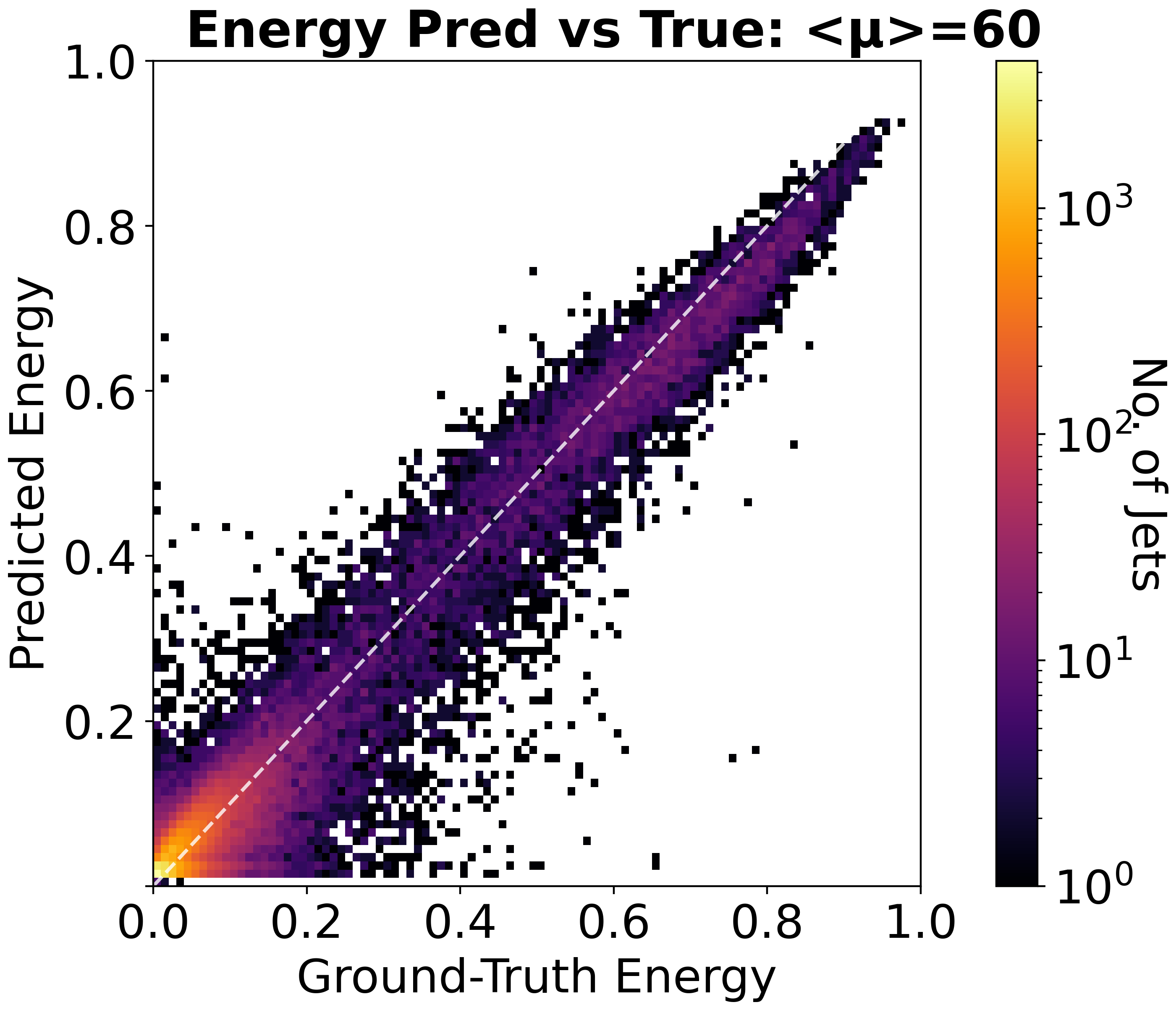}
      \caption{}
      \label{fig:Efrac2d_mu60}
    \end{subfigure}\hfill
    \begin{subfigure}{.20\textwidth}
      \centering
      \includegraphics[width=\linewidth]{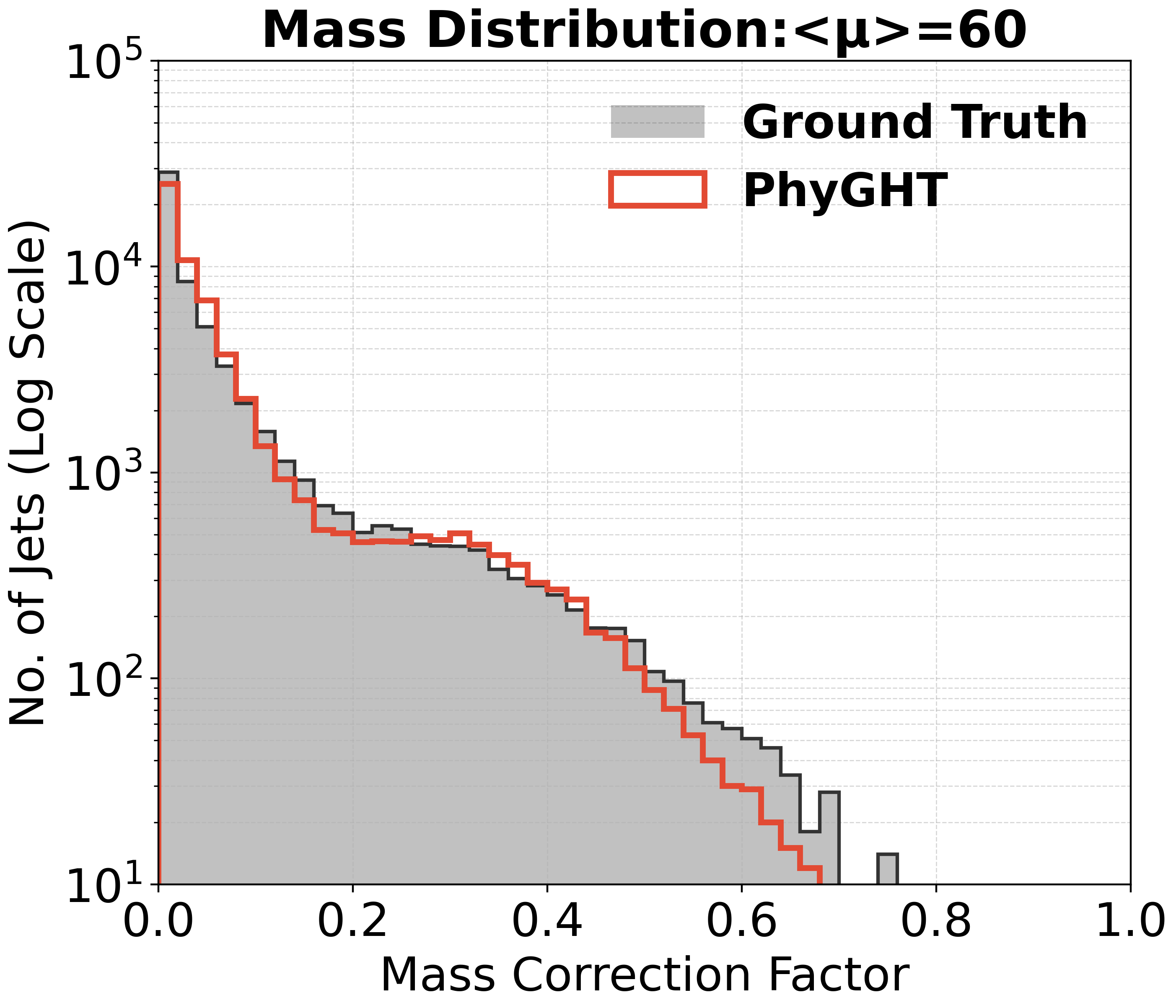}
      \caption{}
      \label{fig:Mfrac1d_mu60}
    \end{subfigure}\hfill
    \begin{subfigure}{.20\textwidth}
      \centering
      \includegraphics[width=\linewidth]{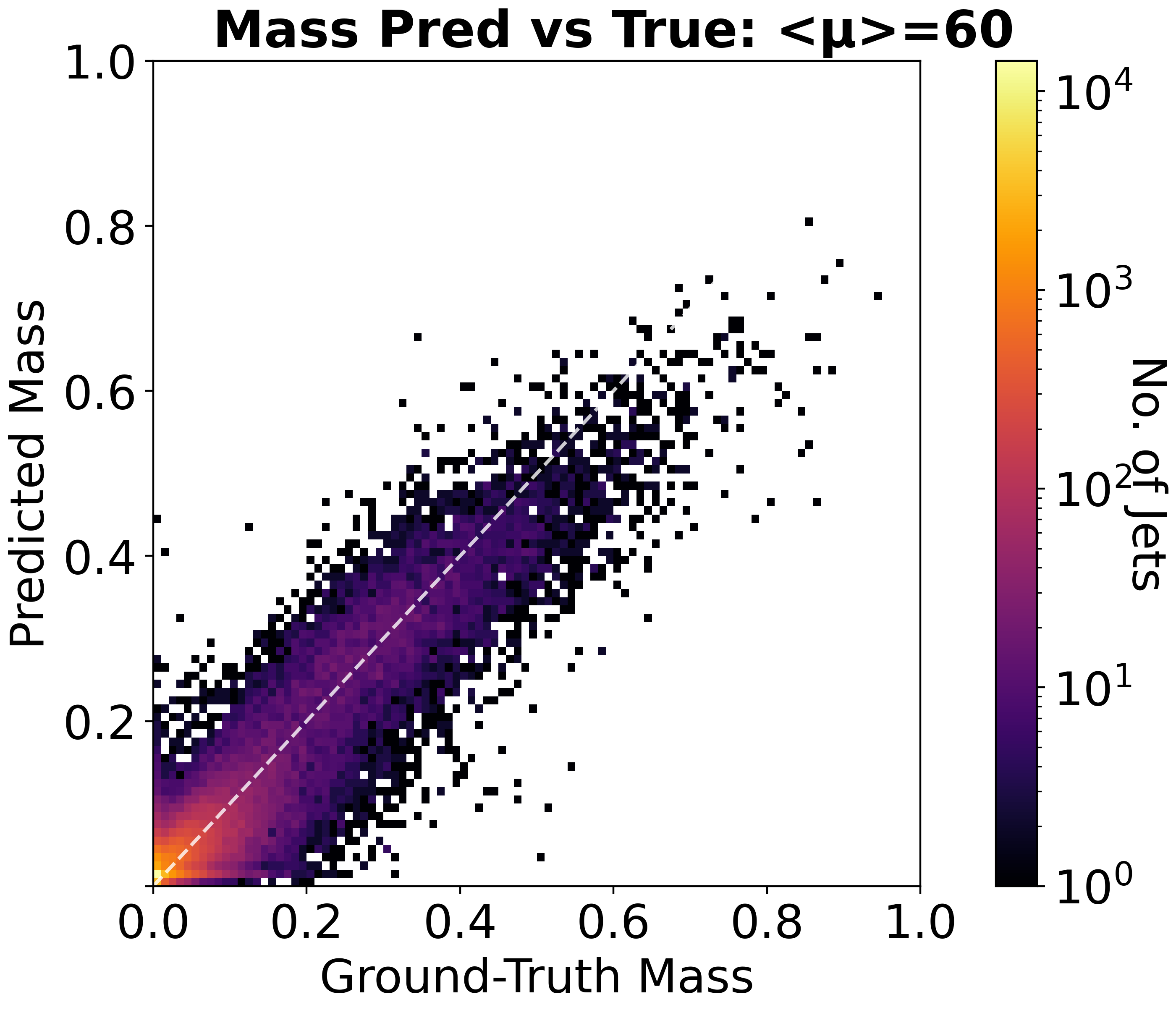}
      \caption{}
      \label{fig:Mfrac2d_mu60}
    \end{subfigure}
    
    % \vspace{1em} % Add space between rows

    % --- Row 2: mu = 200 ---
    \begin{subfigure}{.20\textwidth}
      \centering
      \includegraphics[width=\linewidth]{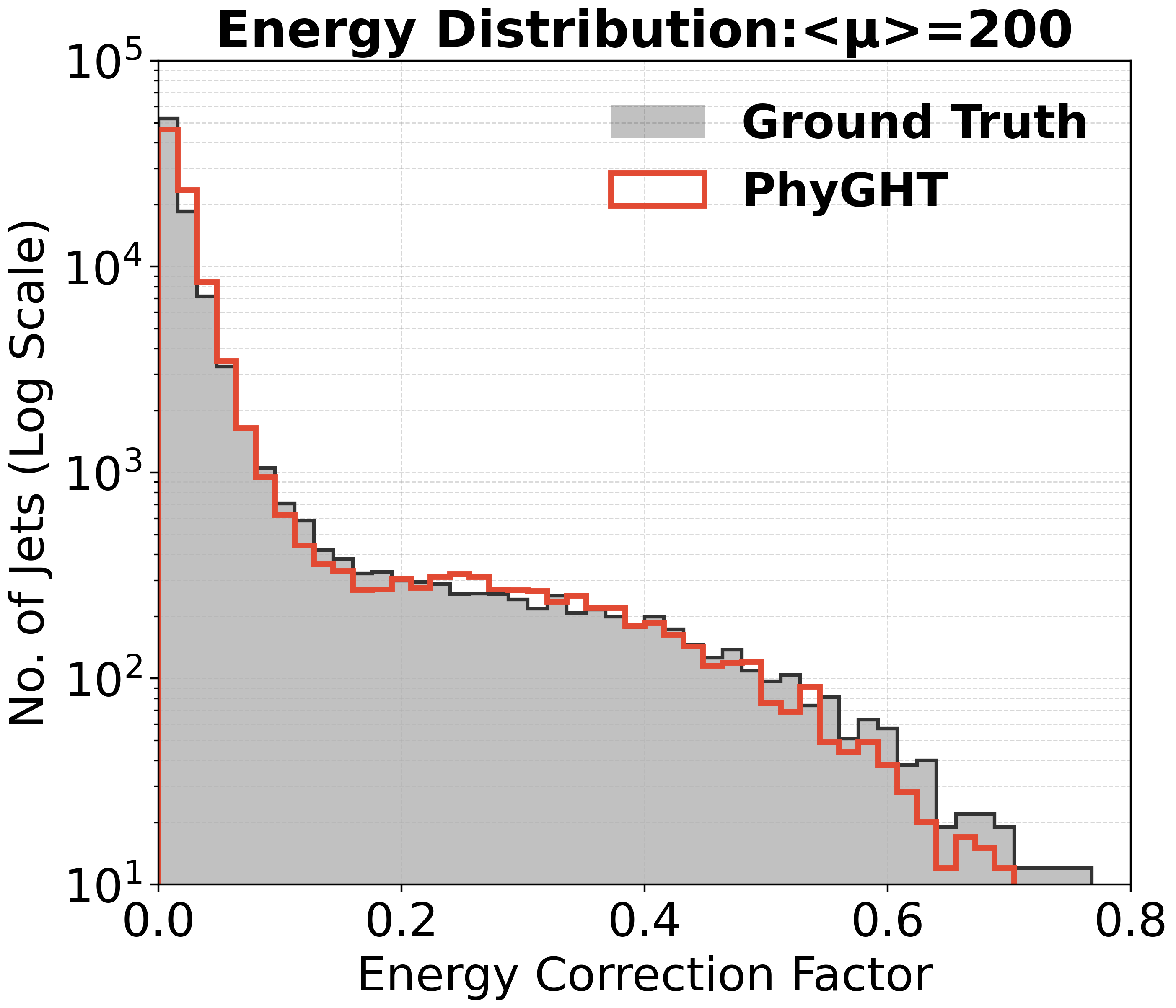}
      \caption{}
      \label{fig:Efrac1d_mu200}
    \end{subfigure}\hfill
    \begin{subfigure}{.20\textwidth}
      \centering
      \includegraphics[width=\linewidth]{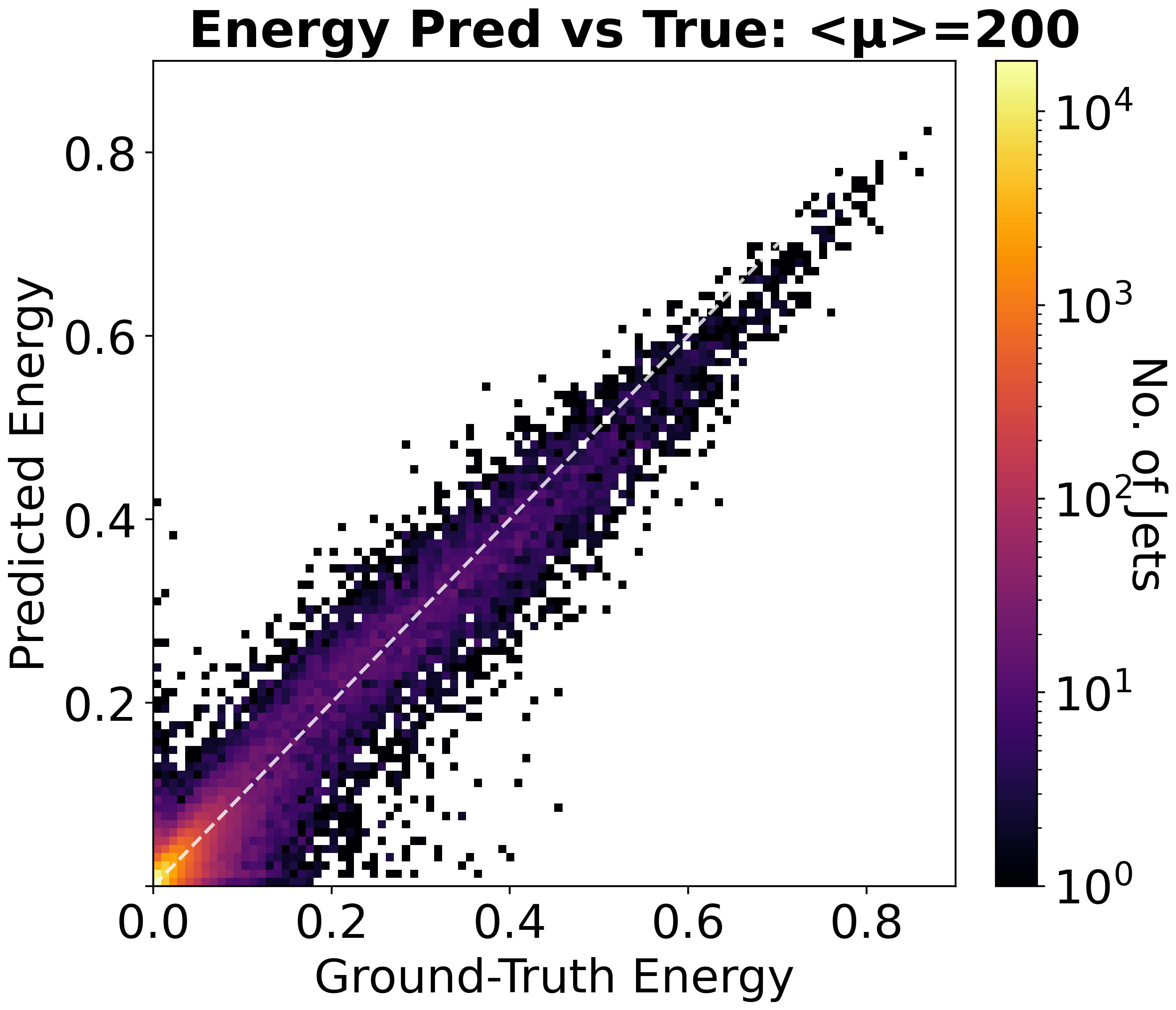}
      \caption{}
      \label{fig:Efrac2d_mu200}
    \end{subfigure}\hfill
    \begin{subfigure}{.20\textwidth}
      \centering
      \includegraphics[width=\linewidth]{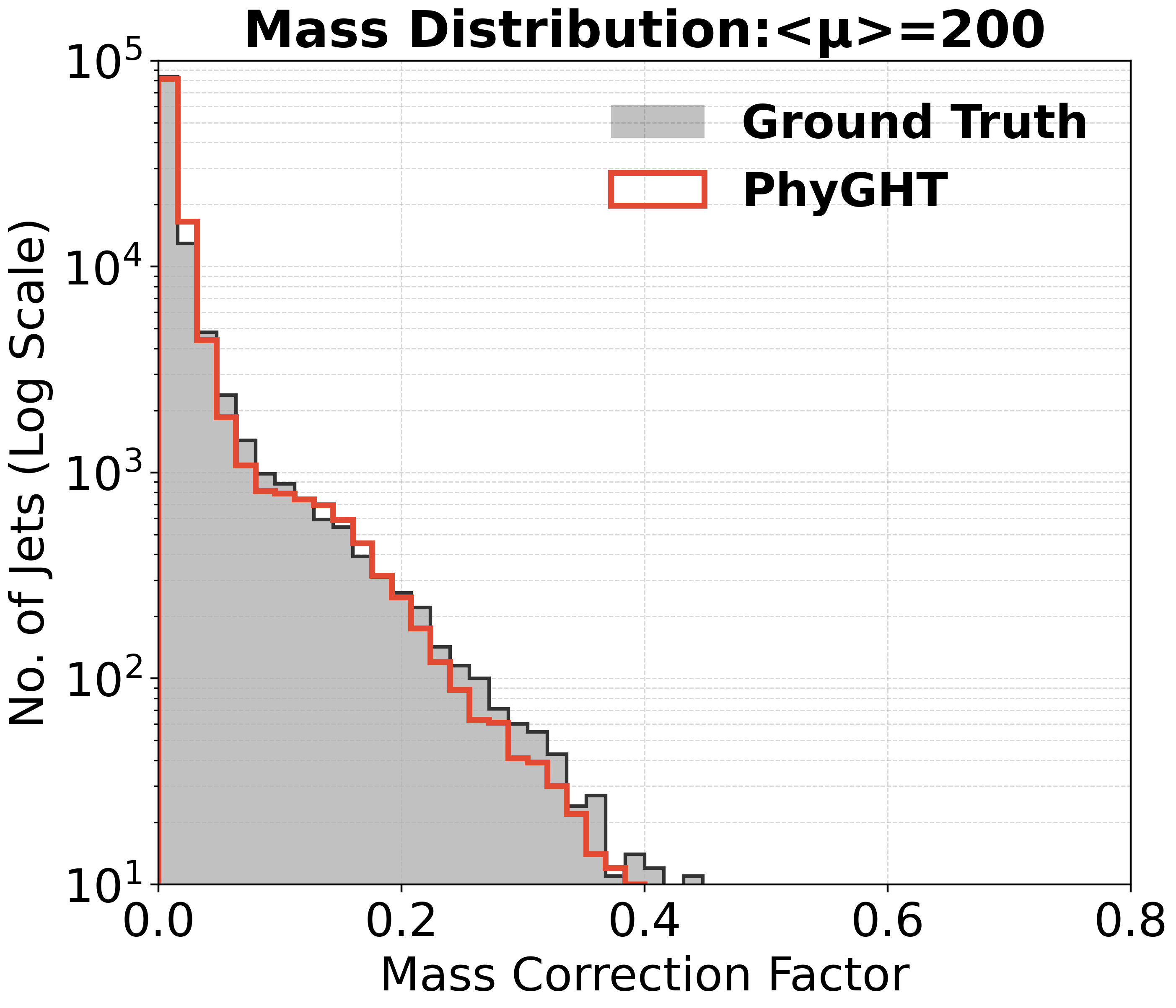}
      \caption{}
      \label{fig:Mfrac1d_mu200}
    \end{subfigure}\hfill
    \begin{subfigure}{.20\textwidth}
      \centering
      \includegraphics[width=\linewidth]{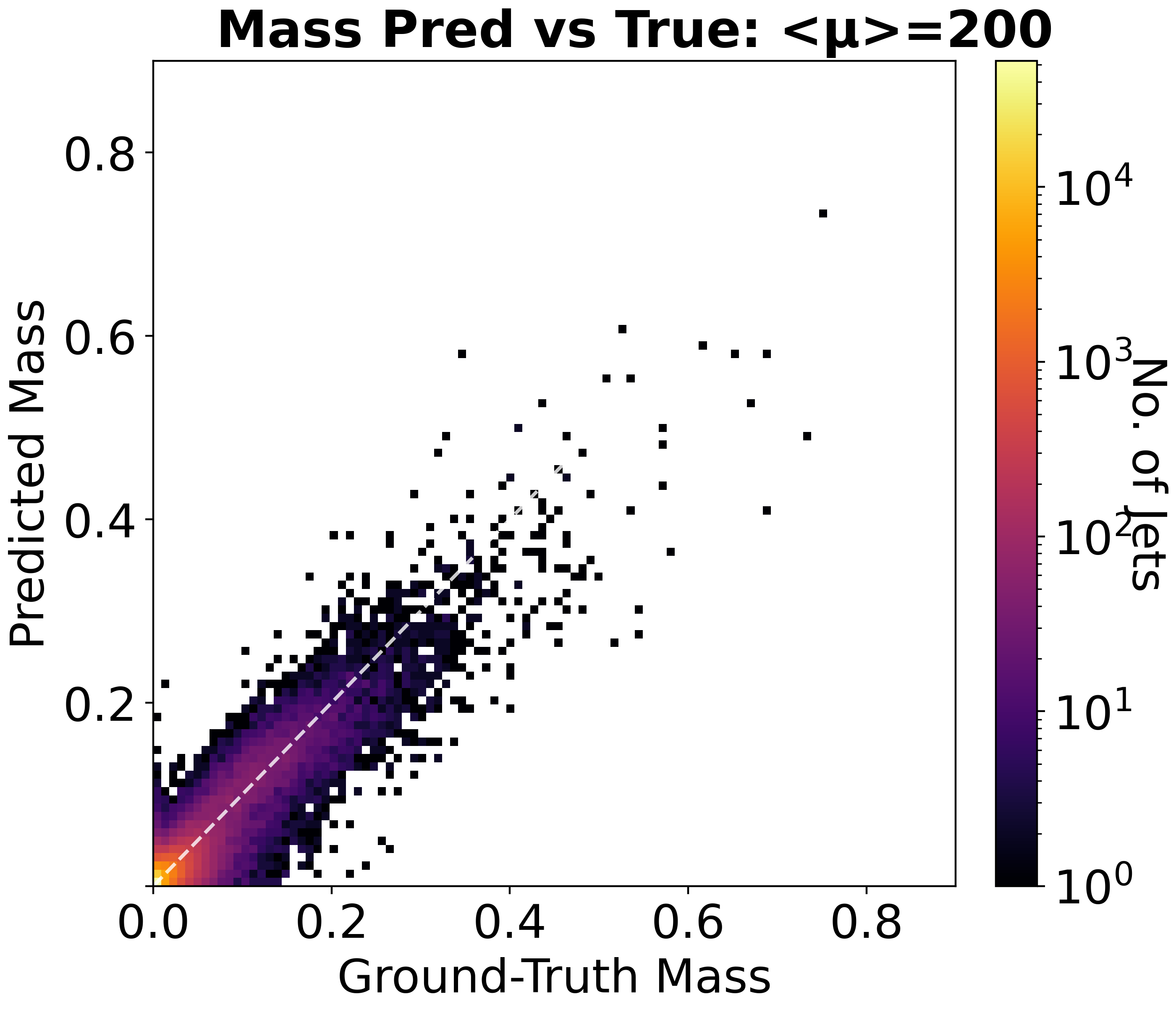}
      \caption{}
      \label{fig:Mfrac2d_mu200}
    \end{subfigure}
    \caption{1D distribution and 2D correlation plots for Energy ($\hat{y}_E$) and Mass ($\hat{y}_M$) at $\langle \mu \rangle=60$ (top row) and $\langle \mu \rangle=200$ (bottom row).}
    \label{fig:RegressionResults}
\end{figure*}

\section{Experiments and Results} 
\label{sec:results}

We evaluate the proposed PhyGHT framework on the basis of the following research questions: 
\textbf{Q1:} How does PhyGHT compare to state-of-the-art baselines in recovering signal observables under extreme pileup conditions? 
\textbf{Q2:} How does the computation efficiency of PhyGHT compare to baseline models for offline reconstruction?
\textbf{Q3:} What is the contribution of each architectural component to the overall model performance and robustness? 
\textbf{Q4:} Does the interpretable PSG gate effectively distinguish signal from background compared to existing physics algorithms?
\textbf{Q5:} Can PhyGHT restore precision for downstream physics tasks, such as the invariant mass resolution of the top quark?

\subsection{Experimental Setup}

\subsubsection{Dataset}
Using $t\bar{t}$ collision dataset detailed in Section~\ref{sec:dataset}, we evaluate performance under two distinct pileup scenarios: standard LHC ($\langle \mu \rangle = 60$) and extreme HL-LHC ($\langle \mu \rangle = 200$). For each scenario, the dataset consisting of 10k events is split into an 80/10/10 proportions for training , validation, and testing, respectively.

\subsubsection{Baselines}
We benchmark PhyGHT against the standard physics algorithm PUPPI \cite{puppi}. To assess geometric deep learning performance, we compare against GNN \cite{gnn}, GAT \cite{gat}, HGNN \cite{hgnn}, and HGAT \cite{hgat}. We also evaluate the sequential Transformer \cite{transformer} to isolate the impact of pure attention mechanisms. Finally, we benchmark against specialized models for high-energy physics, specifically ParticleNet \cite{particlenet} and PUMINet \cite{puminet}. Please refer to Appendix~\ref{ap:baseline} for detailed baseline descriptions.

\subsubsection{Implementation Details}
All experiments were conducted on a single NVIDIA A10 GPU. To ensure a fair comparison, all models were trained for 200 epochs using the AdamW optimizer with a learning rate of $3 \times 10^{-4}$. We utilized a batch size of 16 for $\langle \mu \rangle = 60$, and 4 for $\langle \mu \rangle = 200$. For PhyGHT, we set the nearest neighbor count $k=8$ and the auxiliary loss weight $\lambda_{aux}=0.1$, based on the hyperparameter analysis in Section~\ref{sec:hyperparameter}. Additional hyperparameter details are provided in Appendix~\ref{ap:hyperparameter}.

\begin{figure*}
    \centering
    % --- Energy Resolution (mu60) ---
    \begin{subfigure}{.20524\textwidth}
        \centering
        \includegraphics[width=\linewidth]{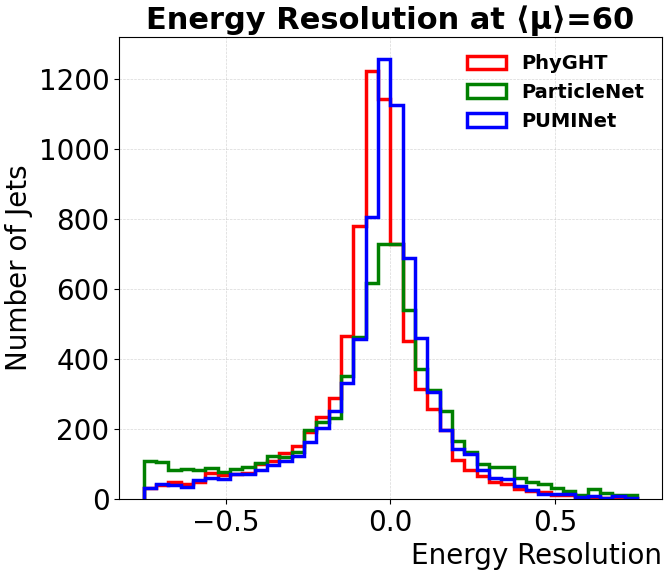}
        \caption{}
        \label{fig:EnergyReso60}
    \end{subfigure}\hfill
    % --- Energy Resolution (mu200) ---
    \begin{subfigure}{.20\textwidth}
        \centering
        \includegraphics[width=\linewidth]{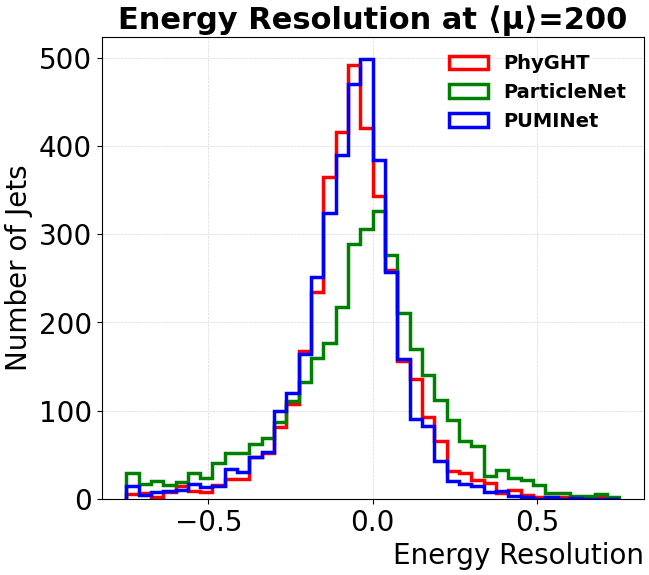}
        \caption{}
        \label{fig:EnergyReso200}
    \end{subfigure}\hfill
    % --- Mass Resolution (mu60) ---
    \begin{subfigure}{.20\textwidth}
        \centering
        \includegraphics[width=\linewidth]{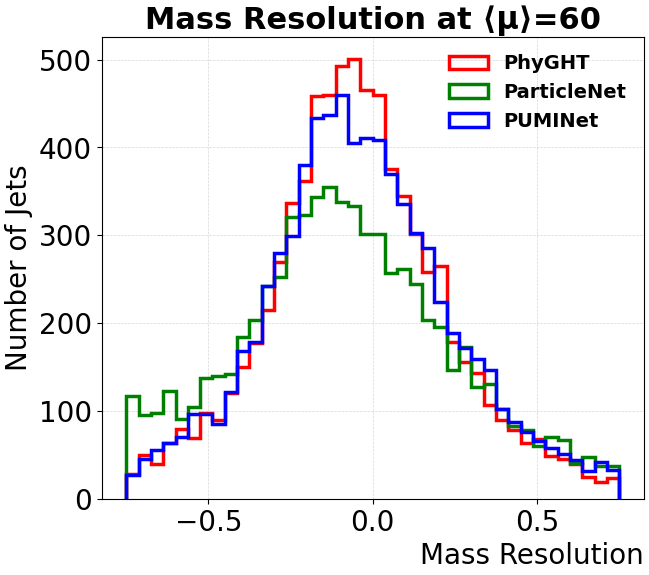}
        \caption{}
        \label{fig:MassReso60}
    \end{subfigure}\hfill
    % --- Mass Resolution (mu200) ---
    \begin{subfigure}{.20\textwidth}
        \centering
        \includegraphics[width=\linewidth]{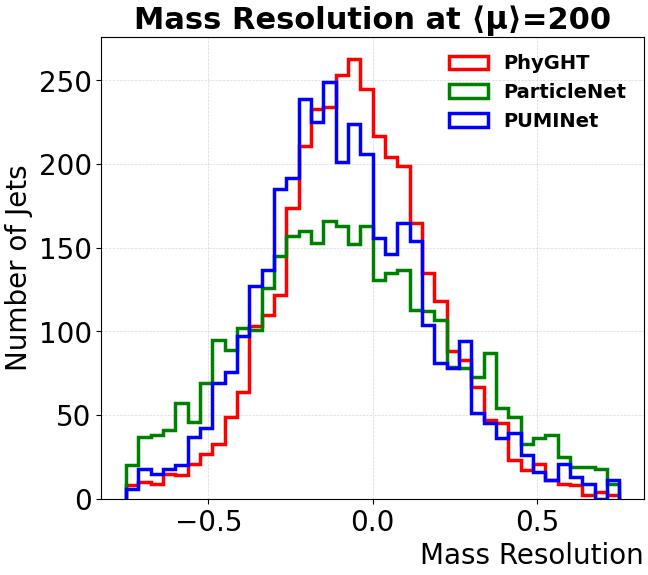}
        \caption{}
        \label{fig:MassReso200}
    \end{subfigure}\hfill
    
    \caption{Comparison of Energy and Mass Resolutions of PhyGHT with baselines at $\langle \mu \rangle=60$ and  $\langle \mu \rangle=200$.}
    \label{fig:Resolution}
\end{figure*}
% \begin{figure}
%     \centering
%     \begin{subfigure}{.23\textwidth}
%       \centering
%       \includegraphics[width=\linewidth]{figures/lukes_upload/Energy_Resolution_mu200.png}
%       \caption{}
%       \label{fig:MassReso60}
%     \end{subfigure}\hfill
%     \begin{subfigure}{.23\textwidth}
%       \centering
%       \includegraphics[width=\linewidth]{figures/lukes_upload/Mass_Resolution_mu200.png}
%       \caption{}
%       \label{fig:MassReso200}
%     \end{subfigure}\hfill
%     \caption{Comparison of Energy and Mass Resolutions}
%     \label{fig:Resolution}
% \end{figure}

\subsection{Results}
We evaluate the model using the Coefficient of Determination ($R^2$). Results with additional metrics (\textit{MAE, MSE, RMSE}) are available in Appendix~\ref{ap:results_table}. In all reported results, \textbf{bold} indicates the best performance, while \underline{underlined} values denote the second-best results.

\subsubsection{Reconstruction Accuracy}
As shown in Table~\ref{tab:main_results}, PhyGHT consistently outperforms all baselines across both standard ($\langle\mu\rangle=60$) and extreme ($\langle\mu\rangle=200$) pileup scenarios.
PhyGHT exhibits exceptional precision in predicting the mass correction factor. This gain is driven by our \textit{distance-aware graph attention} of the local block, which explicitly weights particle interactions based on spatial proximity and preserves the angular correlations critical for accurate signal purification.
The regression plots in Figure~\ref{fig:RegressionResults} further demonstrate the superiority of PhyGHT in pileup mitigation. The 1D distributions (Figs.~\ref{fig:Efrac1d_mu60}, \ref{fig:Mfrac1d_mu60}, \ref{fig:Efrac1d_mu200}, \ref{fig:Mfrac1d_mu200}) confirm that PhyGHT closely tracks the ground truth spectrum across the entire jet frequency spectrum, while the 2D correlation densities (Figs.~\ref{fig:Efrac2d_mu60}, \ref{fig:Mfrac2d_mu60}, \ref{fig:Efrac2d_mu200}, \ref{fig:Mfrac2d_mu200}) show a tight diagonal alignment, demonstrating robust per-jet reconstruction even in the high-energy tails.

\begin{table}[b]
\centering
\caption{Comparison of performance, model size, and inference latency across a single event for 1000 runs.}
\label{tab:efficiency}
\resizebox{0.85\linewidth}{!}{%
\begin{tabular}{c|c|cccc} 
\toprule
\multirow{2}{*}{\textbf{$\bm{\langle \mu \rangle}$}} & \multirow{2}{*}{\textbf{Model}} & \textbf{Energy}    & \textbf{Mass}      & \textbf{Params}  & \textbf{Latency}   \\
                                                     &                                 & ($R^2$) $\uparrow$ & ($R^2$) $\uparrow$ & (M) $\downarrow$ & (ms) $\downarrow$  \\ 
\cmidrule(r){1-6}
\multirow{3}{*}{60}                                  & ParticleNet                     & 0.869              & 0.748              & \textbf{0.17}    & 130.1              \\
                                                     & PUMINet                         & \underline{0.934}  & \underline{0.838}  & \underline{0.85} & \underline{12.4}   \\
                                                     & \textbf{PhyGHT}                 & \textbf{0.943}     & \textbf{0.869}     & 0.95             & \textbf{10.0}      \\ 
\midrule
\multirow{3}{*}{200}                                 & ParticleNet                     & 0.853              & 0.693              & \textbf{0.17}    & 353.1              \\
                                                     & PUMINet                         & \underline{0.926}  & \underline{0.805}  & \underline{0.85} & \underline{78.2}   \\
                                                     & \textbf{PhyGHT}                 & \textbf{0.932}     & \textbf{0.836}     & 0.95             & \textbf{40.4}      \\
\bottomrule
\end{tabular}
}
\end{table}

\subsubsection{Resolution Analysis}
To quantify the precision of our reconstruction, we analyze the resolution distributions defined by the relative error $(y_{pred} - y_{true}) / y_{true}$. Figure~\ref{fig:Resolution} compares PhyGHT against the top-performing baselines, PUMINet and ParticleNet. PhyGHT (red curve) exhibits the sharpest peak centered at zero for both energy and mass, indicating minimal bias and the lowest variance among all methods. This high resolution confirms the effectiveness of our model not only in recovering hard scatter energy and mass of each jet, but also in effectively filtering pileup tracks.

\subsubsection{Computational Efficiency}
Table~\ref{tab:efficiency} benchmarks model performance against computational cost. Despite a marginal increase in parameter count, PhyGHT achieves the lowest latency, delivering a \textit{1.9x} and \textit{8.7x} speedup over PUMINet and ParticleNet, respectively, at $\langle\mu\rangle=200$. This efficiency stems from fundamental structural differences. Baselines suffer from compounding overheads: ParticleNet dynamically recomputes neighbor graphs at every layer, while PUMINet repeats expensive quadratic attention three times in sequence. In contrast, PhyGHT computes the local k-NN graph only once and restricts dense global operations to a single block. By handling local and jet-level aggregation with efficient sparse graph layers, our model avoids recalculating the global event context multiple times. This ensures that latency scales favorably with event complexity, enabling high throughput for offline pileup mitigation.

\begin{table}
\centering
\caption{Cross-process generalization performance. Trained on $t\bar{t}$ dataset and evaluated on the unseen di-Higgs dataset.}
\label{tab:cross_process_transfer}
\resizebox{0.65\linewidth}{!}{%
\begin{tabular}{c|ccc} 
\toprule
\textbf{Model} & \textbf{Energy} & \textbf{Mass} & \textbf{Latency} \\
               & ($R^2$) $\uparrow$ & ($R^2$) $\uparrow$ & (ms) $\downarrow$ \\ 
\cmidrule(r){1-4}
ParticleNet     & 0.888             & 0.754             & 348               \\
PUMINet         & \underline{0.938} & \underline{0.831} & \underline{75}    \\
\textbf{PhyGHT} & \textbf{0.946}    & \textbf{0.864}    & \textbf{39}       \\
\bottomrule
\end{tabular}
}
\end{table}

% \begin{table}
% \centering
% \caption{Impact of removing key components from PhyGHT.}
% \label{tab:ablation_components}
% \resizebox{\linewidth}{!}{%
% \begin{tabular}{c|c|cccc} 
% \toprule
% \begin{tabular}[c]{@{}c@{}}\textbf{Metric}\\\(R^2\)\end{tabular} & \textbf{PhyGHT} & \begin{tabular}[c]{@{}c@{}}\textbf{\textit{w/o} }\\\textbf{Local}\end{tabular} & \begin{tabular}[c]{@{}c@{}}\textbf{\textit{w/o} }\\\textbf{Global}\end{tabular} & \begin{tabular}[c]{@{}c@{}}\textbf{\textit{w/o} }\\\textbf{PSG}\end{tabular} & \begin{tabular}[c]{@{}c@{}}\textbf{\textit{w/o} }\\\textbf{Hypergraph}\end{tabular}  \\ 
% \cmidrule(r){1-6}
% Energy                                                           & \textbf{0.943}  & 0.934                                                                          & 0.848~                                                                          & 0.935                                                                        & 0.931                                                                                \\
% Mass                                                             & \textbf{0.869}  & 0.849                                                                          & 0.715~                                                                          & 0.854~                                                                       & 0.851                                                                                \\
% \bottomrule
% \end{tabular}
% }
% \end{table}
\begin{table}
\centering
\caption{Impact of removing key components from PhyGHT.}
\label{tab:ablation_components}
\resizebox{0.85\linewidth}{!}{%
\begin{tabular}{c|c|cccc} 
\toprule
\begin{tabular}[c]{@{}c@{}}\textbf{Metric}\\\(R^2\)\end{tabular} & \textbf{PhyGHT} & \begin{tabular}[c]{@{}c@{}}\textbf{\textit{w/o} }\\\textbf{Local}\end{tabular} & \begin{tabular}[c]{@{}c@{}}\textbf{\textit{w/o} }\\\textbf{Global}\end{tabular} & \begin{tabular}[c]{@{}c@{}}\textbf{\textit{w/o} }\\\textbf{PSG}\end{tabular} & \begin{tabular}[c]{@{}c@{}}\textbf{\textit{w/o} }\\\textbf{Hypergraph}\end{tabular}  \\ 
\cmidrule(r){1-6}
Energy                                                           & \textbf{0.943}  & 0.903                                                                          & 0.848~                                                                          & \underline{0.915}                                                                        & 0.891                                                                                \\
Mass                                                             & \textbf{0.869}  & 0.813                                                                          & 0.715~                                                                          & \underline{0.837}~                                                                       & 0.824                                                                                \\
\bottomrule
\end{tabular}
}
\end{table}

% \begin{table}
% \centering
% \small
% \caption{Impact of Local DA-GAT neighborhood size ($k$) on regression performance. For $k=0$, the local block is removed.}
% \label{tab:ablation_k}
% \resizebox{0.80\linewidth}{!}{%
% \begin{tabular}{c|ccccccc} 
% \toprule
% \multirow{2}{*}{\begin{tabular}[c]{@{}c@{}}\textbf{Metric }\\$R^2$\end{tabular}} & \multicolumn{7}{c}{\textbf{Neighbors ($\bm{k}$)}}  \\
%                                                                                     & 0 & 2 & 4 & 8 & 16 & 20 & 32                       \\ 
% \cmidrule(r){1-8}
% Energy                                                                              & 0.934 & 0.937 & 0.938 & 0.939 & 0.938 & 0.939  & 0.921                        \\
% Mass                                                                                & 0.849 & 0.856 & 0.858 & 0.861 & 0.860 & 0.860  & 0.829                        \\
% \bottomrule
% \end{tabular}
% }
% \end{table}
\begin{table}
\centering
\small
\caption{Impact of Local DA-GAT neighborhood size ($k$) on regression performance. For $k=0$, the local block is removed.}
\label{tab:ablation_k}
\resizebox{0.85\linewidth}{!}{%
\begin{tabular}{c|ccccccc} 
\toprule
\multirow{2}{*}{\begin{tabular}[c]{@{}c@{}}\textbf{Metric }\\$R^2$\end{tabular}} & \multicolumn{7}{c}{\textbf{Neighbors ($\bm{k}$)}}  \\
                                                                                    & 0 & 2 & 4 & 8 & 16 & 20 & 32                       \\ 
\cmidrule(r){1-8}
Energy                                                                              & 0.903 & 0.932 & \underline{0.938} & \textbf{0.943} & 0.934 & 0.928  & 0.921                        \\
Mass                                                                                & 0.813 & 0.851 & 0.856 & \textbf{0.869} & \underline{0.858} & 0.853  & 0.849                        \\
\bottomrule
\end{tabular}
}
\end{table}

% \begin{table}[t]
% \centering
% \small
% \caption{Impact of the auxiliary classification loss weight ($\lambda_{aux}$) on overall regression performance.}
% \label{tab:ablation_aux}
% \resizebox{0.80\linewidth}{!}{%
% \begin{tabular}{c|cccccc} 
% \toprule
% \multirow{2}{*}{\begin{tabular}[c]{@{}c@{}}\textbf{Metric }\\$R^2$\end{tabular}} & \multicolumn{6}{c}{\textbf{Auxiliary Weight ($\bm{\lambda_{aux}}$)}} \\
%                                                                                  & 0.0 & 0.1 & 0.25 & 0.5 & 0.75 & 1.0 \\ 
% \cmidrule(r){1-7}
% Energy                                                                           & 0.932  & 0.941  & 0.940 & 0.938  & 0.935  & 0.939  \\
% Mass                                                                             & 0.849  & 0.865  & 0.865 & 0.862  & 0.859  & 0.861  \\
% \bottomrule
% \end{tabular}
% }
% \end{table}
\begin{table}
\centering
\small
\caption{Impact of the auxiliary classification loss weight ($\lambda_{aux}$) on overall regression performance.}
\label{tab:ablation_aux}
\resizebox{0.85\linewidth}{!}{%
\begin{tabular}{c|cccccc} 
\toprule
\multirow{2}{*}{\begin{tabular}[c]{@{}c@{}}\textbf{Metric }\\$R^2$\end{tabular}} & \multicolumn{6}{c}{\textbf{Auxiliary Weight ($\bm{\lambda_{aux}}$)}} \\
                                                                                 & 0.0 & 0.1 & 0.25 & 0.5 & 0.75 & 1.0 \\ 
\cmidrule(r){1-7}
Energy                                                                           & 0.915  & \textbf{0.943}  & \underline{0.940} & 0.938  & 0.935  & 0.931  \\
Mass                                                                             & 0.837  & \textbf{0.869}  & \underline{0.865} & 0.862  & 0.859  & 0.855  \\
\bottomrule
\end{tabular}
}
\end{table}

\subsection{Ablation Study}

\subsubsection{Cross-Process Transfer}
To evaluate whether PhyGHT generalizes beyond the semi-leptonic $t\bar{t}$ topology used for training, we test all models directly on an unseen di-Higgs sample at $\langle\mu\rangle=200$ without retraining or fine-tuning. This setting probes whether the learned representation is specific to a single decay topology or captures a more general pileup-suppression mechanism. As shown in Table~\ref{tab:cross_process_transfer}, PhyGHT achieves the best energy and mass reconstruction performance while also maintaining the lowest inference latency. These results suggest that PhyGHT does not merely overfit to the original $t\bar{t}$ process, but transfers effectively to a distinct physics process with different jet topology.

\subsubsection{Impact of Architectural Components} 
Table~\ref{tab:ablation_components} highlights the contribution of each module to the model's predictive power by removing key components from PhyGHT. The \textit{Global Context} block is the most critical component, as its removal isolates the model from the event-wide context required to estimate background pileup density. Among the structural components, the \textit{Hypergraph Aggregation} is the most influential for energy accuracy, enabling the network to selectively gather signal-dominant tracks while ignoring background fluctuations. In contrast, the \textit{Local Geometric} block is indispensable for mass recovery, as preserving local angular correlations is essential for accurately reconstructing the jet's invariant mass. Finally, \textit{PSG} provides a crucial layer of refinement by explicitly filtering noisy constituents before they reach the aggregation stage. Implementation details are in Appendix~\ref{ap:component}.

\subsubsection{Sensitivity to Hyperparameters}
\label{sec:hyperparameter}
We examine the model's sensitivity to key hyperparameters in Tables~\ref{tab:ablation_k} and~\ref{tab:ablation_aux}. The local neighborhood size exhibits a clear optimum at $k=8$, representing the ideal trade-off where the model captures necessary structural correlations without integrating distant pileup noise. For the gating mechanism, a low auxiliary weight of $\lambda_{aux}=0.1$ proves most effective as it scales the auxiliary classification loss to match the magnitude of the primary regression objective. This prevents the gating task from dominating the optimization process.

\begin{figure}
    \centering
    % \begin{subfigure}{.2069\textwidth}
    \begin{subfigure}{.23069\textwidth}
      \centering
      \includegraphics[width=\linewidth]{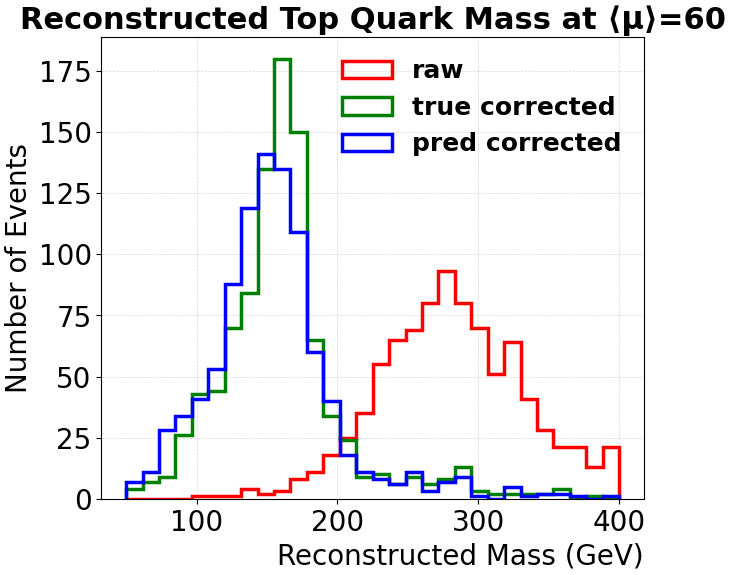}
      \caption{}
      \label{fig:TopReco60}
    \end{subfigure}\hfill
    \begin{subfigure}{.23069\textwidth}
      \centering
      \includegraphics[width=\linewidth]{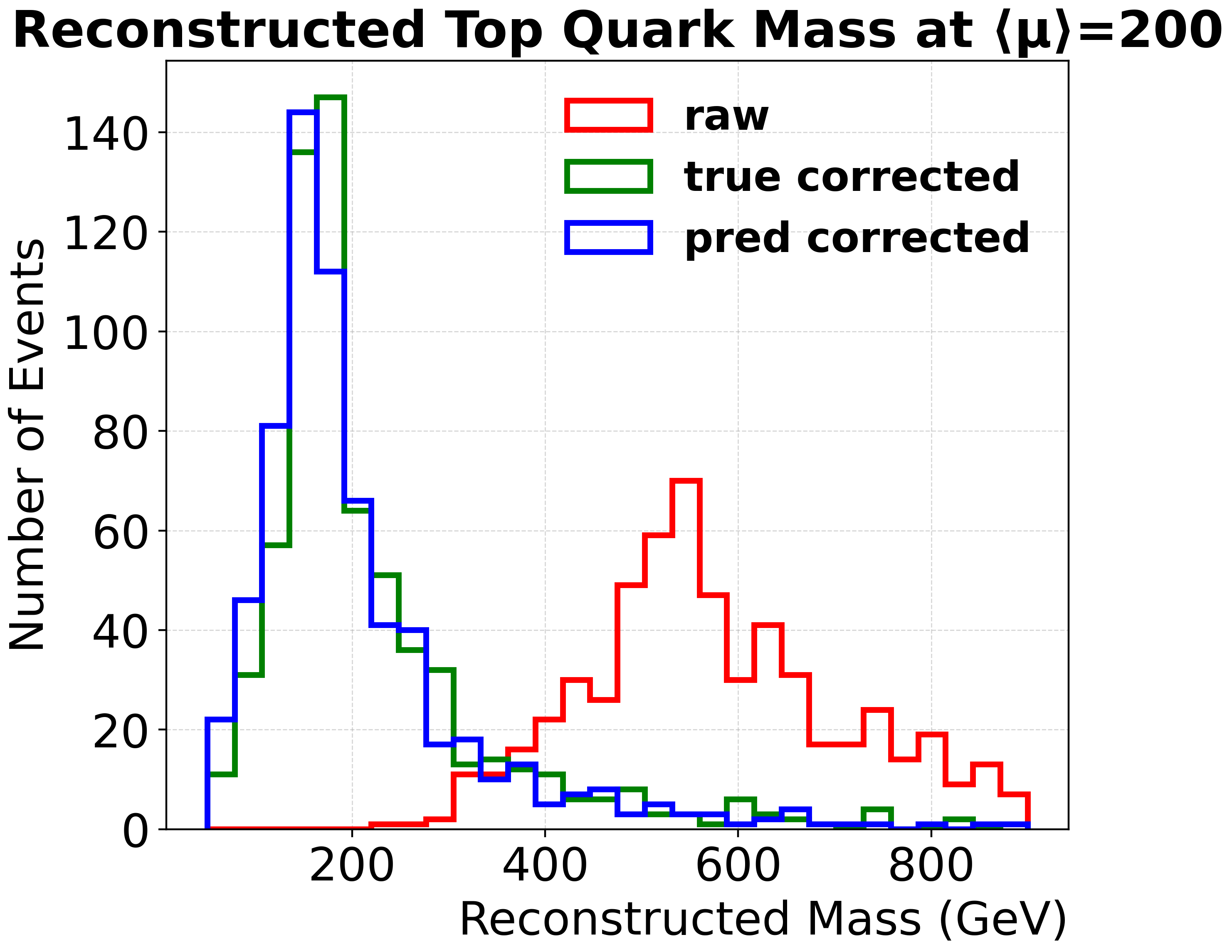}
      \caption{}
      \label{fig:TopReco200}
    \end{subfigure}\hfill
    \caption{Mass Reconstruction after PhyGHT correction}
\end{figure}

\subsection{Physics Study}

\subsubsection{Top Quark Mass Reconstruction}
Figure~\ref{fig:TopReco60} and Figure~\ref{fig:TopReco200} illustrates the reconstruction of the top quark invariant mass, a standard benchmark for practical physics analysis. Heavily distorted by pileup, the uncorrected response (red) shows a mass resonance that is shifted with substantial broadening. PhyGHT successfully mitigates pileup contamination (blue) and nearly matches the ground truth mass resonance (green). Since the top quark mass resonance using PhyGHT's predictions aligns in good agreement with ground truth, we demonstrate that the model can be used in real world physics analysis. Additionally, we have reduced the pileup mitigation task to two simple correction factors which overall simplifies the pileup mitigation workflow compared to existing methods. The detailed methodology for selecting candidate jets and reconstructing the invariant mass is provided in Appendix~\ref{app:mass_reco}.

\begin{table}
\centering
\caption{Impact of generator and detector shifts on regression performance at $\langle\mu\rangle=200$ without retraining.}
\label{tab:robustness_generator_detector}
\resizebox{\linewidth}{!}{%
\begin{tabular}{c|cc|cc|cc} 
\toprule
\textbf{Test Setting} 
& \multicolumn{2}{c|}{\textbf{ParticleNet}} 
& \multicolumn{2}{c|}{\textbf{PUMINet}} 
& \multicolumn{2}{c}{\textbf{PhyGHT}} \\ 
\cmidrule(r){2-3} \cmidrule(r){4-5} \cmidrule(l){6-7}
& \textbf{Energy} & \textbf{Mass} 
& \textbf{Energy} & \textbf{Mass} 
& \textbf{Energy} & \textbf{Mass} \\
& ($R^2$) $\uparrow$ & ($R^2$) $\uparrow$
& ($R^2$) $\uparrow$ & ($R^2$) $\uparrow$
& ($R^2$) $\uparrow$ & ($R^2$) $\uparrow$ \\
\midrule
Pythia Gen. 
& 0.873 & 0.772 
& \underline{0.933} & \underline{0.849} 
& \textbf{0.943} & \textbf{0.880} \\

Inflated $d_0,z_0$ 
& 0.869 & 0.748 
& \underline{0.934} & \underline{0.839} 
& \textbf{0.944} & \textbf{0.869} \\

Both 
& 0.873 & 0.772 
& \underline{0.933} & \underline{0.849} 
& \textbf{0.943} & \textbf{0.880} \\
\bottomrule
\end{tabular}
}
\end{table}

\subsubsection{Robustness to Generator and Detector Shifts}
To evaluate the robustness of PhyGHT under more realistic deployment conditions, we test the trained models under three perturbed settings at $\langle\mu\rangle=200$. First, we evaluate on events generated with an alternative Pythia-based setup to test sensitivity to Monte Carlo generator shift. Second, we inflate the detector resolution noise for the impact-parameter features $d_0$ and $z_0$ by 50\% to test robustness to geometric measurement uncertainty. Third, we combine both perturbations to evaluate the joint effect of generator and detector mismatch. As shown in Table~\ref{tab:robustness_generator_detector}, PhyGHT remains stable across all three settings and consistently outperforms ParticleNet and PUMINet for both energy and mass correction. This suggests that the distance-aware local geometry used by PhyGHT does not rely on overly precise simulated coordinates and remains effective under reasonable perturbations to the generator and detector.

% \subsubsection{Pileup Suppression Efficacy}
% Figure~\ref{fig:ROC} evaluates track-level classification performance under high pileup conditions with $\langle\mu\rangle=200$. PhyGHT achieves a near-perfect AUC of \textit{0.996}, demonstrating that PSG effectively discriminates signal from pileup and enables precise downstream mass and energy corrections. To contextualize this performance, we benchmark against PUPPI \cite{puppi}, a widely-adopted statistical pileup mitigation technique that reweights particles based on their hard-scatter probability. As shown in Table~\ref{main_results}, jet energy and mass predictions using PUPPI weights yield substantially lower $R^2$ values compared to PhyGHT, indicating that modern machine learning approaches offer significant improvements in pileup mitigation accuracy.
\subsubsection{Pileup Suppression Efficacy}
\begin{figure}[!t]
   \centering
    \includegraphics[width=0.49\linewidth]{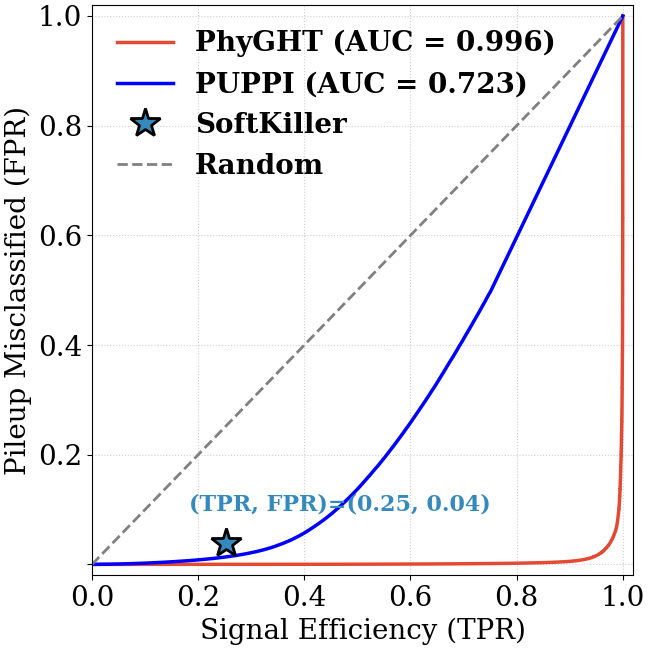}
    \caption{ROC Curve for PSG Track Classification at $\langle \mu \rangle=200$}
   \label{fig:ROC}
\end{figure}

Figure~\ref{fig:ROC} evaluates track-level classification performance under high pileup conditions ($\langle\mu\rangle=200$). PhyGHT achieves a near-perfect ROC curve, demonstrating effective discrimination between signal and pileup. To contextualize this, we benchmark against PUPPI~\cite{puppi}, a standard statistical technique that reweights particles based on their likelihood of originating from a hard scatter. We also compare against SoftKiller~\cite{softkiller}, a geometric approach that applies a median-based $p_T$ cut within grid patches to remove background. As shown in the plot, PhyGHT significantly outperforms both baselines: it maintains high signal efficiency where PUPPI struggles to separate classes, and it avoids the signal loss inherent to SoftKiller's hard fixed cuts. PUPPI's inability to cleanly separate signal from noise directly leads to the poor energy and mass reconstruction seen in Table~\ref{tab:main_results}. This confirms that our physics-guided architecture offers significant improvements in pile-up mitigation accuracy.

\begin{table}
\centering
\caption{Agreement b/w PSG and hypergraph attention. High and low denote the top and bottom 20\% of tracks within each jet. Percentages are calculated over all constituent-track instances after per-jet ranking.}
\label{tab:psg_attention_agreement}
\resizebox{0.80\linewidth}{!}{%
\begin{tabular}{c|c} 
\toprule
\textbf{PSG-Hypergraph Attention Relation} & \textbf{Percentage} \\ 
\midrule
High PSG \& High Attention & 26.8\% \\
Low PSG \& Low Attention & 24.2\% \\
High PSG \& Low Attention & 18.1\% \\
Low PSG \& High Attention & 0.01\% \\
\bottomrule
\end{tabular}
}
\end{table}

\begin{table}[!b]
\centering
\caption{Effect of masking tracks ranked by PSG.}
\label{tab:psg_masking}
\resizebox{0.55\linewidth}{!}{%
\begin{tabular}{c|cc} 
\toprule
\textbf{Masking Setting} & \textbf{Energy} & \textbf{Mass} \\
                         & ($R^2$) $\uparrow$ & ($R^2$) $\uparrow$ \\ 
\midrule
No masking      & 0.932 & 0.836 \\
Mask top 20\%   & 0.459 & 0.093 \\
Mask bottom 20\% & 0.932 & 0.836 \\
\bottomrule
\end{tabular}
}
\end{table}

\subsubsection{PSG Interpretability and Attention Consistency}
To examine the interpretability of PSG, we analyze whether its track-level signal scores are consistent with the downstream hypergraph attention mechanism. For each jet, we rank constituent tracks independently using their PSG scores and hypergraph attention weights. We then define high-score and low-score tracks as the top and bottom 20\% of tracks within each jet, respectively. As shown in Table~\ref{tab:psg_attention_agreement}, the conflicting case where PSG assigns a low signal score but hypergraph attention assigns high importance occurs in only 0.01\% of constituent-track instances. This indicates that tracks identified as pileup-like by PSG are almost never emphasized during jet-level aggregation. In contrast, the high-PSG and low-attention case is not necessarily contradictory: PSG estimates whether a track is signal-like, while hypergraph attention determines whether that signal-like track is useful for correcting a specific jet

We further test whether PSG identifies tracks that are important for the final regression task. As shown in Table~\ref{tab:psg_masking}, masking the top 20\% PSG-ranked tracks sharply degrades both energy and mass reconstruction, whereas masking the bottom 20\% leaves performance unchanged. This confirms that PSG is not only accurate as a track-level classifier, but also identifies the tracks that materially contribute to the final jet correction.

\section{Discussion}
\label{sec:discussion}

The experimental results demonstrate that explicitly modeling the hierarchical topology of particle collision improves pileup mitigation. The \textit{Global Context} block is crucial because it enables the model to analyze the entire event to estimate background density. The \textit{Distance-Aware GAT} significantly improves jet purification by preserving essential angular correlations within the jet's local context. \textit{PSG} effectively filters noise, ensuring the model processes only physically relevant information. This is complemented by the \textit{Hypergraph Aggregation} mechanism, which allows the model to dynamically attend to signal-dominant tracks while ignoring pileup fluctuations. This highlights a fundamental difference in motivation from the original hypergraph neural networks, which were designed to model multi-modal connectivity (e.g., in social media)~\cite{hgnn}; instead, we leverage this topology to enforce the physical hierarchy of particle interactions.
% By restricting dense global operations to a single block and computing the local graph only once, PhyGHT achieves a significant reduction in inference latency compared to the baselines. Ultimately, integrating these components yields a robust framework that predicts precise correction factors capable of restoring physical observables such as the top quark mass. 
Beyond high-energy physics, this framework offers a generalizable solution for any domain that requires separating dense, local signal clusters from global environmental noise. 
% The ability to resolve the trade-off between local geometric precision and global context awareness makes 
This architecture may be applicable to any heterogeneous graph problem where signal and background are topologically distinct, such as denoising 3D point clouds in autonomous driving or detecting anomalous communities in large-scale social networks.

\section{Conclusion}
\label{sec:conclusion}

In this work, we proposed PhyGHT, a physics-guided hypergraph transformer designed to recover hard-scatter observables at the HL-LHC. By leveraging a hierarchical architecture that synergizes global context with local geometry, the model surpasses existing state-of-the-art baselines in pileup suppression. Beyond predictive performance, the model achieves significantly lower inference latency, demonstrating its suitability for practical physics analysis, such as jet reconstruction. Additionally, we introduced a novel simulated dataset of top quark pair production in high-pileup environments to rigorously benchmark these capabilities. We open-source our data to bridge a crucial gap between machine learning and high-energy physics, fostering interdisciplinary collaboration in order to deliver AI solutions to the frontier of science.

\section{Limitations and Ethical Considerations}
\label{sec:limitations}

This study establishes PhyGHT's performance using simulations that closely mimic the HL-LHC environment. Our next step is to integrate the model into the ATLAS software framework to process real collision data with full detector effects. We also plan to extend this architecture to other complex physics signatures, such as di-Higgs production and $W/Z$ boson decays. Regarding ethical considerations, this work relies solely on simulated particle physics data and involves no human subjects or conceivable societal risks.

\begin{acks}
This work is supported by the U.S. Department of Energy (DoE) grant \emph{DE-SC0024669}.
\end{acks}
\newpage
\bibliographystyle{ACM-Reference-Format}
\balance
\bibliography{main}

\newpage
\appendix
\section*{Appendix}

\section{Baseline Descriptions}
\label{ap:baseline}

\subsection{Graph Neural Network (GNN) \cite{gnn}}
To evaluate the efficacy of standard message-passing frameworks, we implement a baseline based on the \textit{GraphSAGE} architecture. We model the collision event as a heterogeneous graph containing two distinct node types: tracks and jets. The graph connectivity is defined by three specific edge sets: (1) \textit{local geometric edges} connecting tracks to their spatial neighbors ($\Delta R < 0.4$) to encode local particle density; (2) \textit{hierarchical edges} linking tracks to their constituent jets based on clustering history, enabling bidirectional information flow; and (3) \textit{global context edges} connecting jets to other nearby jets ($\Delta R < 0.8$). We employ \texttt{SAGEConv} layers to perform mean aggregation across these relations, progressively updating the jet node embeddings to regress the final energy and mass correction factors.

\subsection{Graph Attention Network (GAT) \cite{gat}}
To assess the impact of dynamic feature weighting versus static aggregation, we implement a baseline based on the \textit{GATv2} architecture. This model employs the same heterogeneous graph topology as the GNN baseline, using the same node features and edge sets. In contrast to the static mean pooling of \textit{GraphSAGE}, we employ \texttt{GATv2Conv} layers to compute learnable attention coefficients for every edge. This allows the network to assign adaptive weights to local neighbors, constituent tracks, and surrounding jets, dynamically prioritizing informative connections during the message-passing phase before regressing the final correction factors.

\subsection{Hypergraph Neural Network (HGNN) \cite{hgnn}}
To determine if simply grouping tracks into jets is sufficient without auxiliary spatial connections, we implement a baseline based on the \textit{HGNN} architecture. In this model, we treat jets as \textit{hyperedges} that connect variable-sized sets of track \textit{nodes}. Unlike the GNN and GAT baselines, we discard all pairwise geometric edges, relying entirely on the bipartite structure defined by the jet clustering history. The network employs \texttt{HypergraphConv} layers to propagate information using fixed weights determined simply by the number of connected tracks, rather than learnable attention. After this exchange, the updated track features are averaged and concatenated with the original raw jet features to form the final embedding used for regression.

\subsection{Hypergraph Attention Network (HGAT) \cite{hgat}}
To evaluate the benefit of dynamic feature weighting within the bipartite structure, we implement a baseline based on the \textit{HGAT} architecture. Similar to the HGNN, this model represents the event strictly as tracks connected to jet hyperedges, ignoring spatial neighbor connections. However, instead of the fixed averaging used in HGNN, we employ attention-enabled \texttt{HypergraphConv} layers. By explicitly utilizing the jet features as context, the network calculates a learnable attention score for every track-jet pair. This allows the model to dynamically down-weight pileup tracks and prioritize signal constituents during the aggregation. Finally, these refined track representations are concatenated with the original raw jet features to form the final embedding used for regression.

\subsection{ParticleNet \cite{particlenet}}
To benchmark against a standard model in high-energy physics, we implement \textit{ParticleNet}. This architecture was originally introduced for jet tagging tasks—identifying the type of particle that initiated a jet by treating the jet as an unordered \textit{cloud} of particles. It uses \textit{Dynamic Graph CNNs}, where the model continually finds new neighbors for each track based on learned features rather than fixed physical positions. To adapt this model for our pileup mitigation task, we modified the final output stage. Instead of combining all tracks into a single classification score for the whole event, we group the refined track features back into their specific parent jets. We then average these features to predict the energy and mass correction factors for each individual jet.

\subsection{Transformers \cite{transformer}}
To evaluate the capability of pure self-attention mechanisms to learn event topology without explicit structural bias, we implement a baseline based on the standard \textit{Transformer} architecture. In this approach, we flatten the hierarchical event structure into a single sequence. We project both tracks and jets into a shared latent space, distinguishing them via learnable type embeddings. These tokens are concatenated to form a unified input sequence containing all jets and tracks in the event. We employ a standard Transformer Encoder to process this sequence, enabling every token to attend to every other token via global self-attention. Finally, we slice out the updated embeddings corresponding to the jet tokens to regress the energy and mass correction factors.

\subsection{PUMINet \cite{puminet}}
To compare against a state-of-the-art model specifically designed for pileup mitigation, we implement \textit{PUMINet}. Unlike the standard Transformer, which flattens the event, PUMINet preserves the structure where tracks belong to specific jets. The model processes the event through stacked blocks, each performing a specific sequence of updates. First, tracks use \textit{local self-attention} to exchange information only with other tracks inside the same jet, effectively learning the jet's internal shape. These refined track features are then averaged and combined with the jet's features. Next, the tracks use \textit{global self-attention} to look at every other track in the event, capturing the overall pileup density. Finally, the jets use \textit{cross-attention} to look at this global track context, allowing them to adjust their energy and mass predictions based on the surrounding event noise.

\subsection{PUPPI \cite{puppi}}
The PUPPI algorithm is implemented by initializing track 4-vectors and applying cuts on particles with transverse momentum ($p_T< 1$ GeV), $\eta>4.0$, and neutral particles. Tracks are paired within specific
$\Delta R$ ranges (0.02 to 0.3) to calculate local shape parameters $\alpha_i$ based on the momentum-weighted distances of neighboring tracks. Using truth labels to identify pileup contributions, we construct a $\chi^2$ metric from the median and RMS of pileup $\alpha$ values, which is then converted to PUPPI weights via the cumulative $\chi^2$ distribution function. These weights are applied to reweight jet constituents, allowing calculation of predicted energy and mass fractions that can be validated against true pileup labels through $R^2$ scores and ROC curves. Performance degrades significantly at $\langle\mu\rangle=200$ compared to the original PUPPI paper at $\langle\mu\rangle=80$, likely because our hard scatter events lack generator-level filtering and appear more pileup-like. See validation plots in Appendix \ref{ap:PUPPI_Validation}

\begin{table*}
    \centering
    \caption{Detailed regression metrics for Energy ($\hat{y}_E$) and Mass ($\hat{y}_M$) correction factors across standard ($\langle \mu \rangle = 60$) and high-pileup ($\langle \mu \rangle = 200$) scenarios. Best results are \textbf{bolded}, second-best are \underline{underlined}.}
    \label{tab:additional_metrics}
    \resizebox{0.85\textwidth}{!}{
    \begin{tabular}{c|l|cccc|cccc}
    \toprule
    \multirow{2}{*}{\textbf{$\bm{\langle \mu \rangle}$}} & \multirow{2}{*}{\textbf{Model}} & \multicolumn{4}{c|}{\textit{Energy ($\hat{y}_E$)}} & \multicolumn{4}{c}{\textit{Mass ($\hat{y}_M$)}} \\
     & & \textbf{MSE} $\downarrow$ & \textbf{RMSE} $\downarrow$ & \textbf{MAE} $\downarrow$ & \textbf{$R^2$} $\uparrow$ & \textbf{MSE} $\downarrow$ & \textbf{RMSE} $\downarrow$ & \textbf{MAE} $\downarrow$ & \textbf{$R^2$} $\uparrow$ \\
    \midrule
    \multirow{3}{*}{\shortstack{60}} 
    & ParticleNet & 0.0044 & 0.0664 & 0.0429 & 0.8692 & 0.0030 & 0.0549 & 0.0366 & 0.7483 \\
    & PUMINet & \underline{0.0022} & \underline{0.0472} & \underline{0.0308} & \underline{0.9341} & \underline{0.0019} & \underline{0.0440} & \underline{0.0289} & \underline{0.8389} \\
    & \textbf{PhyGHT (Ours)} & \textbf{0.0019} & \textbf{0.0436} & \textbf{0.0281} & \textbf{0.9436} & \textbf{0.0016} & \textbf{0.0396} & \textbf{0.0257} & \textbf{0.8692} \\
    \midrule
    \multirow{3}{*}{\shortstack{200}} 
    & ParticleNet & 0.0009 & 0.0305 & 0.0174 & 0.8536 & 0.0005 & 0.0212 & 0.0125 & 0.6935 \\
    & PUMINet & \underline{0.0005} & \underline{0.0216} & \underline{0.0124} & \underline{0.9265} & \underline{0.0003} & \underline{0.0169} & \underline{0.0097} & \underline{0.8052} \\
    & \textbf{PhyGHT (Ours)} & \textbf{0.0004} & \textbf{0.0207} & \textbf{0.0122} & \textbf{0.9322} & \textbf{0.0002} & \textbf{0.0155} & \textbf{0.0093} & \textbf{0.8364} \\
    \bottomrule
    \end{tabular}
    }
\end{table*}

\section{Hyperparameter Settings}
\label{ap:hyperparameter}

To ensure a fair comparison, we utilized a consistent set of hyperparameters across all baselines and the proposed \textit{PhyGHT} model. All experiments were conducted on a single NVIDIA A10 GPU using the \textit{AdamW} optimizer with a fixed learning rate of $3 \times 10^{-4}$ and a random seed of 42 for reproducibility. We trained all models for 200 epochs, using a batch size of 16 for the standard pileup scenario ($\langle \mu \rangle = 60$) and 4 for the high-pileup scenario ($\langle \mu \rangle = 200$) to accommodate memory constraints. For all architectures, we set the hidden dimension to $d=128$, the number of attention heads to 4, the network depth to $L=3$ layers, and the dropout rate to 0.1. For the PhyGHT model specifically, we set the nearest-neighbor count for local graph construction to $k=8$ and the auxiliary loss weight for the gating mechanism to $\lambda_{aux}=0.1$.

\section{Extended Performance Metrics}
\label{ap:results_table}

Table~\ref{tab:additional_metrics} presents a comprehensive evaluation with MSE, RMSE, MAE, and $R^2$ for the proposed PhyGHT architecture compared to the two most competitive baselines: ParticleNet and PUMINet. Results are reported for both the standard ($\langle \mu \rangle = 60$) and high-luminosity ($\langle \mu \rangle = 200$) pileup scenarios. We observe that PhyGHT outperforms all baselines across all metrics in both pileup scenarios.

\section{Top Quark Mass Reconstruction Process}
\label{app:mass_reco}
To reconstruct the top quark mass, we first apply the predicted corrections to each jet by rescaling the mass and energy according to the predicted fractions. We then define a corrected jet vector with a new momentum magnitude $|\vec{p}|=\sqrt{E^2-m^2}$ and transverse momentum $p_T=|\vec{p}|\text{sech}(\eta)$. This results in a corrected Lorentz 4-vector where the direction remains unchanged, but the kinematics are scaled to reflect only the hard-scatter contributions.

Using truth labels, we trace the parton shower history via a depth-first search algorithm to identify final-state particles originating from the $b$-quark and $W$-boson of the top quark decay. We then select three candidate jets: the two containing the highest fraction of tracks from the $W$-boson are designated as $W_1$ and $W_2$, while the jet with the most particles from the $b$-quark is designated as $B$. The top quark 4-vector is reconstructed by summing these candidates, $\vec{T} = \vec{W}_1+\vec{W}_2+\vec{B}$, and the final invariant mass is calculated using the energy-momentum relation $m=\sqrt{E^2-|\vec{p}|^2}$.

\begin{figure}[!b]
    \centering
    \begin{subfigure}{.23\textwidth}
      \centering
      \includegraphics[width=\linewidth]{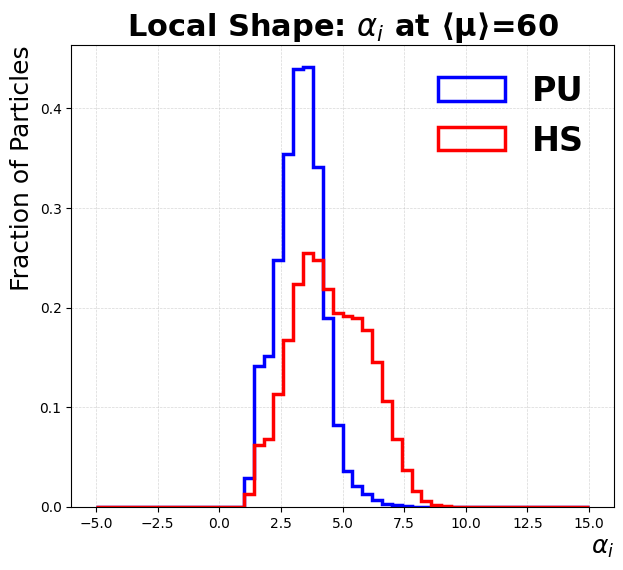}
      \label{fig:alpha_i_mu60}
    \end{subfigure}\hfill
    \begin{subfigure}{.23\textwidth}
      \centering
      \includegraphics[width=\linewidth]{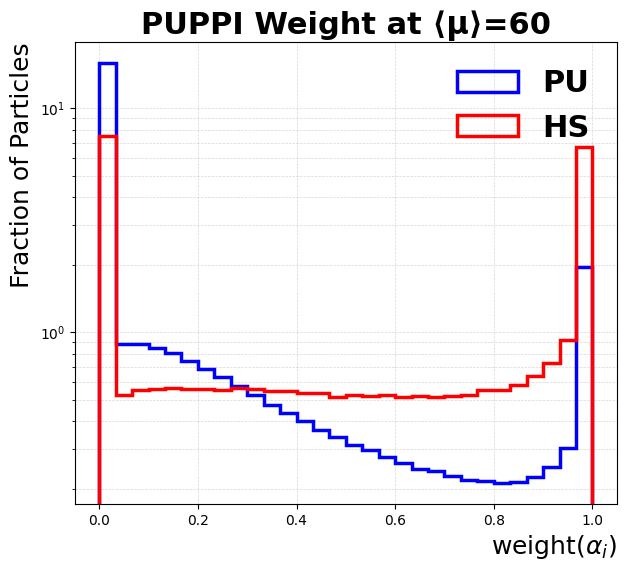}
      \label{fig:PUPPI_weights_mu60}
    \end{subfigure}\hfill
    \begin{subfigure}{.23\textwidth}
      \centering
      \includegraphics[width=\linewidth]{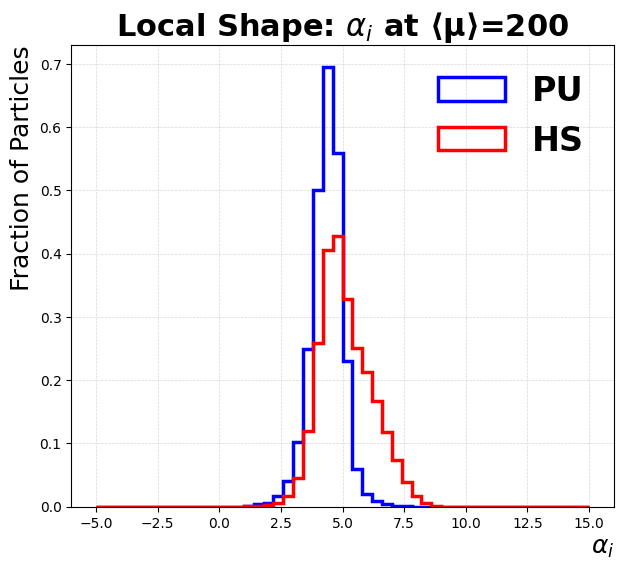}
      \label{fig:alpha_i_mu200}
    \end{subfigure}\hfill
    \begin{subfigure}{.23\textwidth}
      \centering
      \includegraphics[width=\linewidth]{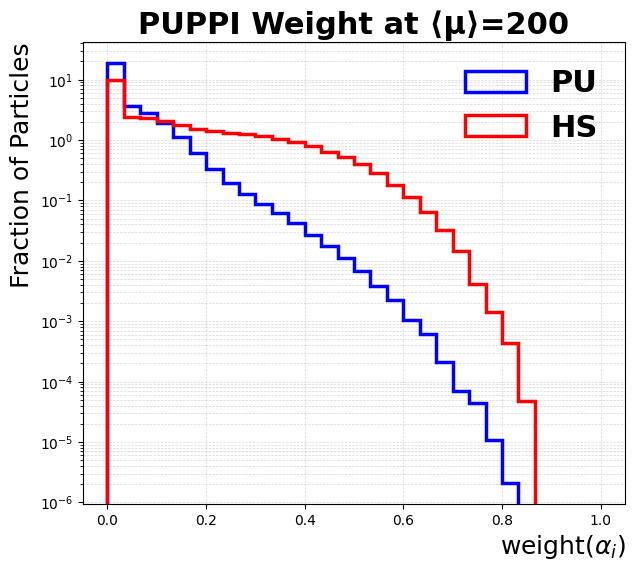}
      \label{fig:PUPPI_weights_mu200}
    \end{subfigure}\hfill
    \caption{Validation Plots for implementation of the PUPPI algorithm}
    \label{fig:PUPPI_valdiation}
\end{figure}

\section{PUPPI Validation}
\label{ap:PUPPI_Validation}
To implement the PUPPI algorithm on our dataset, we first initialize Lorentz 4-vectors of each track using $[p_T,\eta,\phi,m]$ where $m\approx0$ and has charge $q$. Tracks with $p_T< 1GeV$, $\eta>4.0$, and $q=0$ are cut from the dataset. Using awkward library in python, we find all possible pairs of tracks, $[T_i,T_j]$, and cut all pairs of tracks with $\Delta R(T_i,T_j)>0.3$ and $\Delta R(T_i,T_j)<0.02$. For each $T_i$ in all passing pairs, an $\epsilon$ parameter is calculated where $\epsilon_{ij}=\frac{p_{T}(T_j)}{\Delta R(T_i,T_j)}$. Then for all $T_i$, we calculate the local shape parameter $\alpha_i$, using the following equation on pairs that pass the $\Delta R$ cut.

\begin{equation}
\alpha_i = log \left( \sum_{i} \epsilon_{ij} \right)
\end{equation}

Then we select the $\alpha_i$ originating from pileup using truth labels, and calculate the median, $\alpha_{PU}^{median}$, and RMS, $\sigma_{PU}^{^2}$  originating from pileup. We then construct a $\chi^2$ metric using the following equation where $\mathcal{H}$ is the Heavyside function: 

\begin{equation}
\chi^2=\mathcal{H}(\alpha_i-\alpha_{PU}^{median})\frac{(\alpha_i-\alpha_{PU}^{median})^2}{\sigma_{PU}^{^2}}
\end{equation}

A PUPPI Weight is then constructed using using $F_{\chi^2}$, the cumulative distribution function of the $\chi^2$ distribution with a single degree of freedom:

\begin{equation}
    w_i=F_{\chi^2,NDF=1}(\chi^2_i).
\end{equation}

After each track is assigned a PUPPI weight, we reweight each constituent of each jet accordingly. When can sum over the weighted 4-vectors of the set of tracks to calculate the predicted energy and mass fraction of each jet according to PUPPI weights. Since some particles were cut from the dataset with $p_T<1 GeV$, we also recalculate the energy and mass fractions using true pileup labels for the remaining constituents. From these recalculated values, we can derived an $R^2$ score and ROC curve. Note: at $\langle\mu\rangle=200$ the same cuts were used for PUPPI weights, but the performance sharply dropped as shown in the Figure \ref{fig:PUPPI_valdiation}. Since we do not apply a generator level filter to hard scatter events, our hard scatter appears more pileup-like than the original PUPPI paper at $\langle\mu\rangle=80$.

\section{Ablation Implementation Details}
\label{ap:component}

To understand how much each part of our model contributes to the final performance, we trained four different versions of PhyGHT. In each version, we removed or replaced exactly one component while keeping everything else the same. The specific changes are described below.

\subsection{w/o Global Context}
Here, we removed the \textit{Global Transformer} block entirely. In the original model, this block allows tracks to \textit{attend} to every other track in the event to estimate the overall pileup noise. By removing it, the model is forced to rely solely on local information for each track, without knowing what is happening in the rest of the event.

\subsection{w/o Hypergraph Aggregation}
Here, we replaced our specialized \textit{Jet Attention} mechanism with a simple average. In the full model, the Hypergraph allows the jet to assign different importance weights to its tracks (e.g., focusing on high-energy signal tracks). In this ablation, the model simply takes the average feature of all tracks in the jet, treating them all as equally important.

\subsection{w/o Local Geometric}
In this version, we skipped the \textit{Distance-Aware GAT} layer at the very beginning of the network. Normally, this layer helps tracks understand their immediate neighbors within the jet's dense core. By removing it, the raw input features are passed directly to the global block, preventing the model from explicitly learning local spatial relationships between nearby tracks.

\subsection{w/o PSG (Pileup Suppression Gate)}
We removed the \textit{soft gating} network, which serves as a filter before the final aggregation. In the full model, this component calculates a \textit{signal probability} for each track and suppresses those that appear to be noise. For this ablation, we set this probability to 1.0 for every track, effectively turning off the filter and forcing the model to use all tracks—signal and pileup alike to calculate the jet properties.

\end{document}